\pdfoutput=1

\documentclass[aps,prd,twocolumn,superscriptaddress,preprintnumbers,floatfix]{revtex4-1}

\usepackage{graphicx}
\usepackage{amsmath}
\usepackage{mdwlist}
\usepackage{subcaption}
\usepackage{caption}
\usepackage{siunitx}
\usepackage{placeins}

\newcommand{\beq}{\begin{equation}}
\newcommand{\eeq}{\end{equation}}

\newcommand{\hinj}{h_{\rm inj}}
\newcommand{\hrec}{h_{\rm rec}}
\newcommand{\hmin}{h_{\rm min}}
\newcommand{\fracdetthrsh}{\alpha_{\rm n}}
\newcommand{\sigdetthrsh}[1]{s_{#1}}
\newcommand{\fracbandconf}{\alpha_{\rm b}}

\begin{document}

\preprint{LIGO-P1400169}

% Title and authors
\title{Detecting Beyond-Einstein Polarizations of Continuous Gravitational Waves}

\author{Maximiliano Isi}
\email[]{misi@ligo.caltech.edu}

\author{Alan J. Weinstein}
\email[]{ajw@ligo.caltech.edu}

\author{Carver Mead}
\email[]{carver@caltech.edu}
\affiliation{California Institute of Technology}

\author{Matthew Pitkin}
\email[]{matthew.pitkin@glasgow.ac.uk}
\affiliation{University of Glasgow}

\date{\today}

\begin{abstract}
The direct detection of gravitational waves with the next generation detectors, like Advanced LIGO, provides the opportunity to measure deviations from the predictions of General Relativity. One such departure would be the existence of alternative polarizations. To measure these, we study a single detector measurement of a continuous gravitational wave from a triaxial pulsar source. We develop methods to detect signals of any polarization content and distinguish between them in a model independent way.  We present LIGO S5 sensitivity estimates for 115 pulsars.
\end{abstract}

\maketitle

\section{Introduction}
Since its introduction in 1915, Einstein's theory of General Relativity (GR) has been confirmed by experiment in every occasion \cite{Will2006}. However, GR has not yet been tested with great precision on scales larger than the solar system or for highly dynamical and strong gravitational fields \cite{Turyshev2008}. Those kinds of rapidly changing fields give rise to gravitational waves (GWs)---self propagating stretching and squeezing of spacetime originating in the acceleration of massive objects, like spinning neutron stars with an asymmetry in their moment of inertia (e.g.,~see \cite{Ostriker1969, Shklovskii1970}).

Although GWs are yet to be directly observed, detectors such as the Laser Interferometer Gravitational Wave Observatory (LIGO) expect to do so in the coming years, giving us a chance to probe GR on new grounds \cite{Harry2010, Weinstein2011}. Because GR does not present any adjustable parameters, these tests have the potential to uncover new physics \cite{Will2006}. By the same token, LIGO data could also be used to test alternative theories of gravity that disagree with GR on the properties of GWs.

Furthermore, when looking for a weak signal in noisy LIGO data, certain physical models are used to target the search and are necessary to make any detection possible \cite{Turyshev2008}. Because these are usually based on predictions from GR, assuming an incorrect model could yield a weak detection or no detection at all. Similarly, if GR is not a correct description for highly dynamical gravity, checking for patterns given by alternative models could result in detection where no signal had been seen before.

There exist efforts to test GR by looking at the deviations of the parametrized post--Newtonian coefficients extracted from the inspiral phase of compact binary coalescence events \cite{DelPozzo2011, Li2012, Agathos2014}. Besides this, deviations from GR could be observed in generic GW properties such as polarization, wave propagation speed or parity violation \cite{Will2006, Chatziioannou2012, Yunes2010}. Tests of these properties have been proposed which make use of GW burst search methods \cite{Horava2014}.

In this paper, we present methods to search LIGO--like detector data for continuous GW signals of any polarization mode, not just those allowed by GR. We also compare the relative sensitivity of different model--dependent and independent templates to certain kinds of signals. Furthermore, we provide expected sensitivity curves for GR and non--GR signals, obtained by means of blind searches over LIGO noise (not actual upper limits).

Section \ref{sec:background} provides the background behind GW polarizations and continuous waves, while sections \ref{sec:method} and \ref{sec:analysis} present search methods and the data analysis procedures used to evaluate sensitivity for detection. Results and final remarks are provided in sections \ref{sec:results} \& \ref{sec:conclusions} respectively.

\section{Background} \label{sec:background}
\subsection{Polarizations}

\begin{figure}
\begin{subfigure}[c]{0.33\columnwidth}
\includegraphics[width=\textwidth]{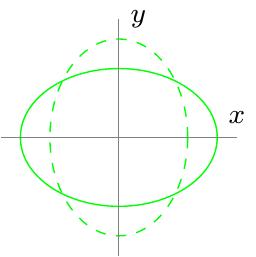}
\end{subfigure}\hfill
\begin{subfigure}[c]{0.33\columnwidth}
\includegraphics[width=\textwidth]{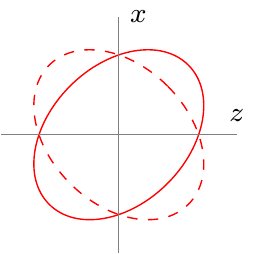}
\end{subfigure}\hfill
\begin{subfigure}[c]{0.33\columnwidth}
\includegraphics[width=\textwidth]{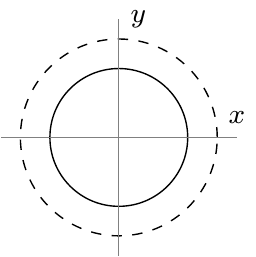}
\end{subfigure}\\
\begin{subfigure}[c]{0.33\columnwidth}
\includegraphics[width=\textwidth]{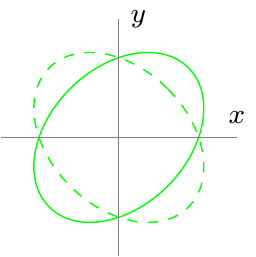}
\end{subfigure}\hfill
\begin{subfigure}[c]{0.33\columnwidth}
\includegraphics[width=\textwidth]{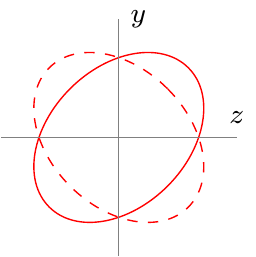}
\end{subfigure}\hfill
\begin{subfigure}[c]{0.33\columnwidth}
\includegraphics[width=\textwidth]{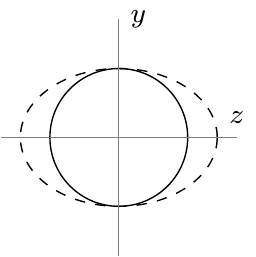}
\end{subfigure}
\caption{Illustration of the effect of different GW polarizations on a ring of test particles. plus (+) and cross ($\times$) tensor modes (green); vector--x (x) and vector--y (y) modes (red);  breathing (b) and longitudinal (l) scalar modes (black). In all of these diagrams the wave propagates in the \emph{z}--direction \cite{Will2006}.\label{fig:circles}}
\end{figure}

Just like electromagnetic waves, GWs can present different kinds of polarizations. Most generally, metric theories of gravity could allow six possible modes: plus ($+$), cross ($\times$), vector x (x), vector y (y), breathing (b) and longitudinal (l). Their effects on a free--falling ring of particles are illustrated in fig.~\ref{fig:circles}.
Transverse GWs ($+$, $\times$ and b) change the distance between particles separated in the plane perpendicular to the direction of propagation (taken to be the $z$-axis). Vector GWs are also transverse; but, because all particles in a plane perpendicular to the direction of propagation are equally accelerated, their relative separation is not changed. Nonetheless, particles farther from the source move at later times, hence varying their position relative to points with both different $x$--$y$ coordinates and different $z$ distance. Finally, longitudinal GWs change the distance between particles separated along the direction of propagation.

Note that, because of their symmetries, the breathing and longitudinal modes are degenerate for LIGO-like interferometric detectors, so it is enough to just consider one of them in the analysis. Also, this study assumes wave frequency and speed remain constant across modes, which restricts the detectable differences between polarizations to amplitude modulations.

In reality, however, GWs might only possess some of those six components: different theories of gravity predict the existence of different polarizations. In fact, due to their symmetries, $+$ and $\times$ are associated with tensor theories, x and y with vector theories, and b and l with scalar theories. In terms of particle physics, this differentiation is also linked to the predicted helicity of the graviton: $\pm$2, $\pm$1 or 0, respectively. Consequently, GR only allows + and $\times$, while scalar--tensor theories also predict the presence of some extra b component whose strength depends on the source \cite{Will2006}. Bolder theories might predict the existence of vector or scalar modes \emph{only}, while still being in agreement with all other non--GW tests.

Four-Vector Gravity (G4v) is one such extreme example \cite{Mead2015}. This vector--based framework claims to reproduce all the predictions of GR, including weak--field tests and total radiated power of GWs. However, this theory differs widely from GR when it comes to gravitational wave polarizations. Thus, one of the only ways to test G4v would be to detect a GW signal composed of x and y modes instead of + and $\times$.

\subsection{Signal} \label{sec:signal}

Because of their persistence, continuous gravitational waves (CGWs) provide the means to study GW polarizations without the need for multiple detectors. For the same reason, continuous signals can be integrated over long periods of time, thus improving the likelihood of detection. Furthermore, these GWs are quasi--sinusoidal and present well--defined frequencies. This allows us to focus on the amplitude modulation, where the polarization information is contained.

CGWs are produced by localized sources with periodic motion, such as binary systems or spinning neutron stars \cite{Zimmermann1979}. Throughout this paper, we target known pulsars (e.g.,~the Crab pulsar) and assume an asymmetry in their moment of inertia (rather than precession of the spin axis or other possible, but less likely, mechanisms) causes them to emit gravitational radiation. A source of this type can generate GWs only at multiples of its rotational frequency $\nu$. In fact, it is expected that most power be radiated at twice this value \cite{Jones2002}. For that reason, we take the GW frequency, $\nu_{\rm gw}$, to be $2\nu$. Moreover, the frequency evolution of these pulsars is well--known thanks to electromagnetic observations, mostly at radio wavelengths but also in gamma-rays.

Simulation of a CGW from a triaxial neutron star is straightforward. The general form of a such signal is:
\begin{equation} \label{eq:cgw}
h(t)= \displaystyle\sum\limits_{p}\ A_{p}(t;\psi|\alpha,\delta,\lambda,\phi,\gamma, \xi)~ h_{p}(t;\iota,h_{0},\phi_{0}, \nu, \dot{\nu}, \ddot{\nu}),
\end{equation}
where, for each polarization \textit{p}, $A_{p}$ is the detector response (antenna pattern) and $h_{p}$ a sinusoidal waveform of frequency $\nu_{\rm gw}=2\nu$. The detector parameters are: $\lambda$, longitude; $\phi$, latitude; $\gamma$, angle of the detector \emph{x}--arm measured from East; and $\xi$, the angle between arms. Values for the LIGO Hanford Observatory (LHO), LIGO Livingston Observatory (LLO) and Virgo (VIR) detectors are presented in table \ref{tab:DetParam}. The source parameters are: $\psi$, the signal polarization angle; $\iota$, the inclination of the pulsar spin axis relative to the observer's line-of-sight; $h_{0}$, an overall amplitude factor; $\phi_{0}$, a phase offset; and $\nu$, the rotational frequency, with $\dot{\nu}$,  $\ddot{\nu}$ its first and second derivatives. Also, $\alpha$ is the right ascension and $\delta$ the declination of the pulsar in celestial coordinates.

Note that the inclination angle $\iota$ is defined as is standard in astronomy, with $\iota=0$ and $\iota=\pi$ respectively meaning that the angular momentum vector of the source points towards and opposite to the observer. The signal polarization angle $\psi$ is related to the position angle of the source, which is in turn defined to be the East angle of the projection of the source's spin axis onto the plane of the sky.

Although there are hundreds of pulsars in the LIGO band, in the majority of cases we lack accurate measurements of their inclination and polarization angles. The few exceptions, presented in table \ref{tab:extrap}, were obtained through the study of the pulsar spin nebula \cite{Ng2008}. This process cannot determine the spin direction, only the orientation of the spin axis. Consequently, even for the best studied pulsars $\psi$ and $\iota$ are only known modulo a reflection: we are unable to distinguish between $\psi$ and $-\psi$ or between $\iota$ and $\pi-\iota$). As will be discussed in section \ref{sec:method}, our ignorance of $\psi$ and $\iota$ must be taken into account when searching for CGWs.

\renewcommand{\arraystretch}{1.2}
\begin{table}[hb]
      \caption{LIGO detectors \cite{Althouse2001}\cite{Allen1996}}
\begin{ruledtabular}
    \begin{tabular}{cccc}
          & LHO   & LLO & VIR \\
\cline{2-4}\\[-10pt]
    Latitude ($\lambda$) & 46.45$ ^{\circ}$ N & 30.56$ ^{\circ}$ N & 43.63$ ^{\circ}$ N \\
    Longitude ($\phi$) & $119.41 ^{\circ}$ W & $90.77 ^{\circ}$ W & $10.5 ^{\circ}$ E \\
    Orientation ($\gamma$) & 125.99$ ^{\circ}$   & 198.0$ ^{\circ}$ & 71.5$ ^{\circ}$ \\ % this is the angle of arm 1 wrt true East (not north as in []. So it's the same as the angle of arm 2 wrt to true north

    \end{tabular}%
\end{ruledtabular}
  \label{tab:DetParam}%
\end{table}%

\begin{table}[hbtp]
\caption{Axis polarization ($\psi$) and inclination ($\iota$) angles for known pulsars \cite{Ng2008}.}
\hrule width \hsize \kern 0.6mm \hrule width \hsize
\begin{minipage} [b]{0.45\linewidth}\centering
    \begin{tabular}{rrr}
     & ${\bf \psi}$ {\scriptsize (deg)} & ${\bf \iota}$ {\scriptsize (deg)} \\
\cline{2-3}\\[-10pt]
     Crab & 124.0 & 61.3\\
    Vela & 130.6 & 63.6 \\
    J1930$+$1852 & 91 & 147\\
     J2229$+$6114 & 103 & 46\\
     B1706$-$44 & 163.6 & 53.3\\
     J2021$+$3651 & 45 & 79\\
     \end{tabular}
     \end{minipage}
    \hfill
~
\hfill
     \begin{minipage} [b]{0.45\linewidth}\centering	
     \begin{tabular}{rrr}
    & $\psi$ {\scriptsize (deg)} & $\iota$ {\scriptsize (deg)} \\
\cline{2-3}\\[-10pt]
     J0205$+$6449 &90.3 & 91.6 \\
     J0537$-$6910 & 131 &92.8 \\
     B0540$-$69 & 144.1 & 92.9\\
     J1124$-$5916 & 16 & 105\\
     B1800$-$21 &  44 & 90 \\
     J1833$-$1034 & 45 & 85.4\\
     \end{tabular}
     \end{minipage}
\hrule width \hsize \kern 0.6mm \hrule width \hsize
\label{tab:extrap}
\end{table}%

\subsubsection{Frequency evolution}
In eq.~(\ref{eq:cgw}), $h_p(t)$ is a sinusoid carrying the frequency modulation of the signal:
\beq \label{eq:h(t)}
h_p(t)=a_p \cos\left(\phi(t) + \phi_p+ \phi_0^{\rm gw}\right)
\eeq
\beq \label{eq:phase}
\phi(t) = 4\pi \left(\nu t_{\rm b} + \frac{1}{2}\dot{\nu}t_{\rm b}^2+\frac{1}{6}\ddot{\nu}t_{\rm b}^3\right) +\phi_0^{\rm em},
\eeq
where $t_{\rm b}$ is the Solar System barycentric arrival time, which is the local arrival time $t$ modulated by the standard R{\o}mer $\Delta_{R}$, Einstein $\Delta_{E}$ and Shapiro $\Delta_{S}$ delays \cite{Edwards2006}:
\beq
t_{\rm b}=t+\Delta_{R}+\Delta_{E}+\Delta_{S}.
\eeq
The leading factor of four in the r.h.s.~of eq.~(\ref{eq:phase}) comes from the substitution $\nu_{\rm gw}=2\nu$. For known pulsars, $\phi_0^{\rm em}$ is the phase of the radio pulse, while $\phi_0^{\rm gw}$ is the phase difference between electromagnetic and gravitational waves. Both factors contribute to an overall phase offset of the signal ($\phi_0^{\rm em} + \phi_0^{\rm gw}$). This is of astrophysical significance since it may provide insights about the relation between EM \& GW radiation and provide information about the physical structure of the source.

The $a_p$ and $\phi_p$ coefficients in eq.~(\ref{eq:h(t)}) respectively encode the relative amplitude and phase of each polarization. These values are determined by the physical model. For instance, GR predicts:
\beq \label{eq:a+}
a_+ = h_0(1+\cos^2\iota)/2 ~,~\phi_+=0,
\eeq
\beq \label{eq:ax}
a_\times= h_0 \cos \iota ~,~\phi_\times=-\pi/2,
\eeq
while $a_{\rm x}=a_{\rm y}=a_{\rm b}=0$. On the other hand, according to G4v \cite{Mead2015}:
\beq \label{eq:axz}
a_{\rm x} = h_0\sin \iota~,~\phi_{\rm x} = -\pi/2,
\eeq
\beq \label{eq:ayz}
a_{\rm y}= h_0 \sin \iota \cos \iota~,~\phi_{\rm x} = 0.
\eeq
while $a_+=a_\times=a_{\rm b}=0$. In both cases, the overall amplitude $h_0$ can be characterized by \cite{Jones2002, Dupuis2005, Mead2015}:
\beq \label{eq:h0}
h_0 = \frac{4 \pi^2 G}{c^4} \frac{I_{zz} \nu^2}{r}\epsilon,
\eeq
where r is the distance to the source, $I_{zz}$ the pulsar's moment of inertia along the principal axis, $\epsilon= (I_{xx}-I_{yy})/I_{zz}$ its equatorial ellipticity  and, as before, $\nu$ is the rotational frequency. Choosing some canonical values,
\beq
h_0\approx 4.2 \times 10^{-26} \frac{I_{zz}}{10^{28}\text{ kg m}^2}\left[\frac{\nu}{100\text{ Hz}}\right]^2 \frac{1\text{ kpc}}{r} \frac{\epsilon}{10^{-6}},
\eeq
it is easy to see that GWs from triaxial neutron stars are expected to be relatively weak \cite{LSC2010}. However, the sensitivity to these waves grows with the observation time because the signal can be integrated over long periods of time \cite{Dupuis2005}.

As indicated in the introduction to this section, we have assumed CGWs are caused by an asymmetry in the moment of inertia of the pulsar. Other mechanisms, such as precession of the spin axis, are expected to produce waves of different strengths and with dominant components at frequencies other than $2\nu$. Furthermore, these effects vary between theories: for instance, in G4v, if the asymmetry is not perpendicular to the rotation axis, there can be a significant $\nu$ component as well as the $2\nu$ component. In those cases, eqs.~(\ref{eq:h(t)}, \ref{eq:h0}) do not hold (e.g.,~see \cite{Jones2002} for precession models).

\subsubsection{Amplitude modulation}
At any given time, GW detectors are not equally sensitive to all polarizations. The response of a detector to a particular polarization $p$ is encoded in a function $A_p(t)$ depending on the relative locations and orientations of the source and detector. As seen from  eq.~(\ref{eq:cgw}), these functions provide the amplitude modulation of the signal.

A GW is best described in an orthogonal coordinate frame defined by wave vectors $({\bf w}_x,~{\bf w}_y,~{\bf w}_z)$, with ${\bf w}_z = {\bf w}_x \times {\bf w}_y$ being the direction of propagation. Furthermore, the orientation of this wave--frame is fixed by requiring that the East angle between ${\bf w}_y$ and the celestial North be $\psi$. In this gauge, the different polarizations act through six orthogonal basis strain tensors \cite{Nishizawa2009, Blaut2012}: %\textbf{[Matt's reference in LAL is \cite{Baut2012}]}:
\begin{align} 
\begin{split} 
e_{jk}^{+}=\begin{pmatrix}
1 & 0 & 0 \\
0 & -1 & 0 \\
0 & 0 & 0
\end{pmatrix}
\end{split},
\begin{split} 
e_{jk}^{\times}=\begin{pmatrix}
0 & 1 & 0 \\
1 & 0 & 0 \\
0 & 0 & 0
\end{pmatrix} 
\end{split}, \tag{2,3} 
\end{align}

\begin{align} 
\begin{split} 
e_{jk}^{\rm x}=\begin{pmatrix}
0 & 0 & 1 \\
0 & 0 & 0 \\
1 & 0 & 0
\end{pmatrix}
\end{split} ,
\begin{split} 
e_{jk}^{\rm y}=\begin{pmatrix}
0 & 0 & 0 \\
0 & 0 & 1 \\
0 & 1 & 0
\end{pmatrix} 
\end{split}, \tag{4,5} 
\end{align}

\begin{align}
\begin{split} 
e_{jk}^{\rm b}=\begin{pmatrix}
1 & 0 & 0 \\
0 & 1 & 0 \\
0 & 0 & 0
\end{pmatrix}
\end{split},
\begin{split} 
e_{jk}^{\rm l}=\sqrt{2} \begin{pmatrix}
0 & 0 & 0 \\
0 & 0 & 0 \\
0 & 0 & 1
\end{pmatrix} 
\end{split}, \tag{6,7} 
\addtocounter {equation} {6}
\end{align}
with $j, k$ indexing $x$, $y$ and $z$ components. These tensors can be written in an equivalent, frame--independent form
\beq
{\bf e}^{+}= {\bf w}_x \otimes {\bf w}_x - {\bf w}_y \otimes {\bf w}_y,
\eeq
\beq
{\bf e}^{\times}=  {\bf w}_x \otimes {\bf w}_y + {\bf w}_y \otimes {\bf w}_x,
\eeq
\beq
{\bf e}^{\rm x}= {\bf w}_x \otimes {\bf w}_z + {\bf w}_z \otimes {\bf w}_x,
\eeq
\beq
{\bf e}^{\rm y}= {\bf w}_y \otimes {\bf w}_z + {\bf w}_z \otimes {\bf w}_y,
\eeq
\beq
{\bf e}^{\rm b}= {\bf w}_x \otimes {\bf w}_x + {\bf w}_y \otimes {\bf w}_y,
\eeq
\beq
{\bf e}^{\rm l}=\sqrt{2}  \left( {\bf w}_z \otimes {\bf w}_z \right).
\eeq

\begin{figure*} [hbtp]
  	\centering
        \begin{subfigure}[b]{0.33\textwidth}
		\centering
                \includegraphics[width=\textwidth]{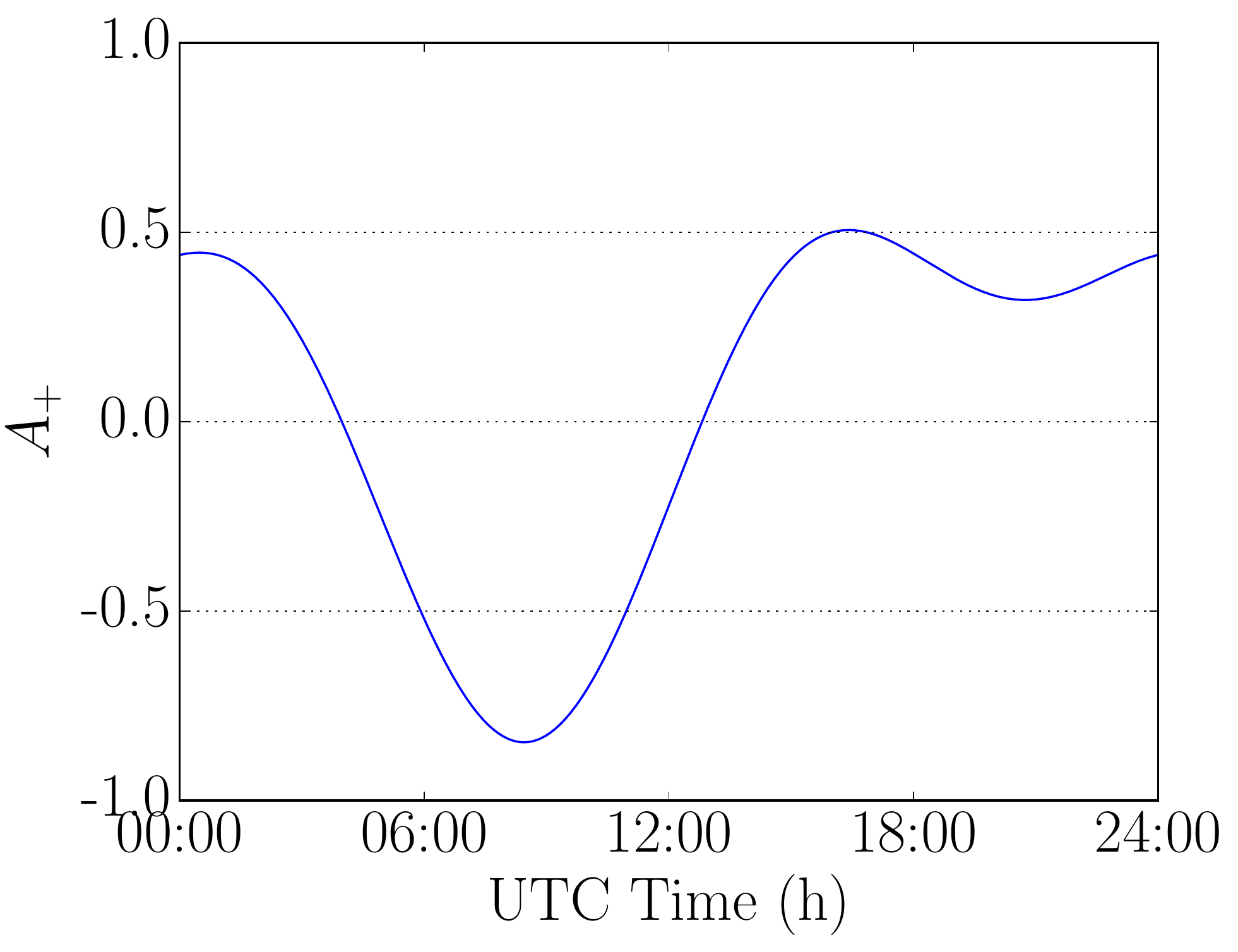}
                \caption{Plus (+)}
                \label{fig:pl}
        \end{subfigure}%
        \hfill
        \begin{subfigure}[b]{0.33\textwidth}
		\centering
                \includegraphics[width=\textwidth]{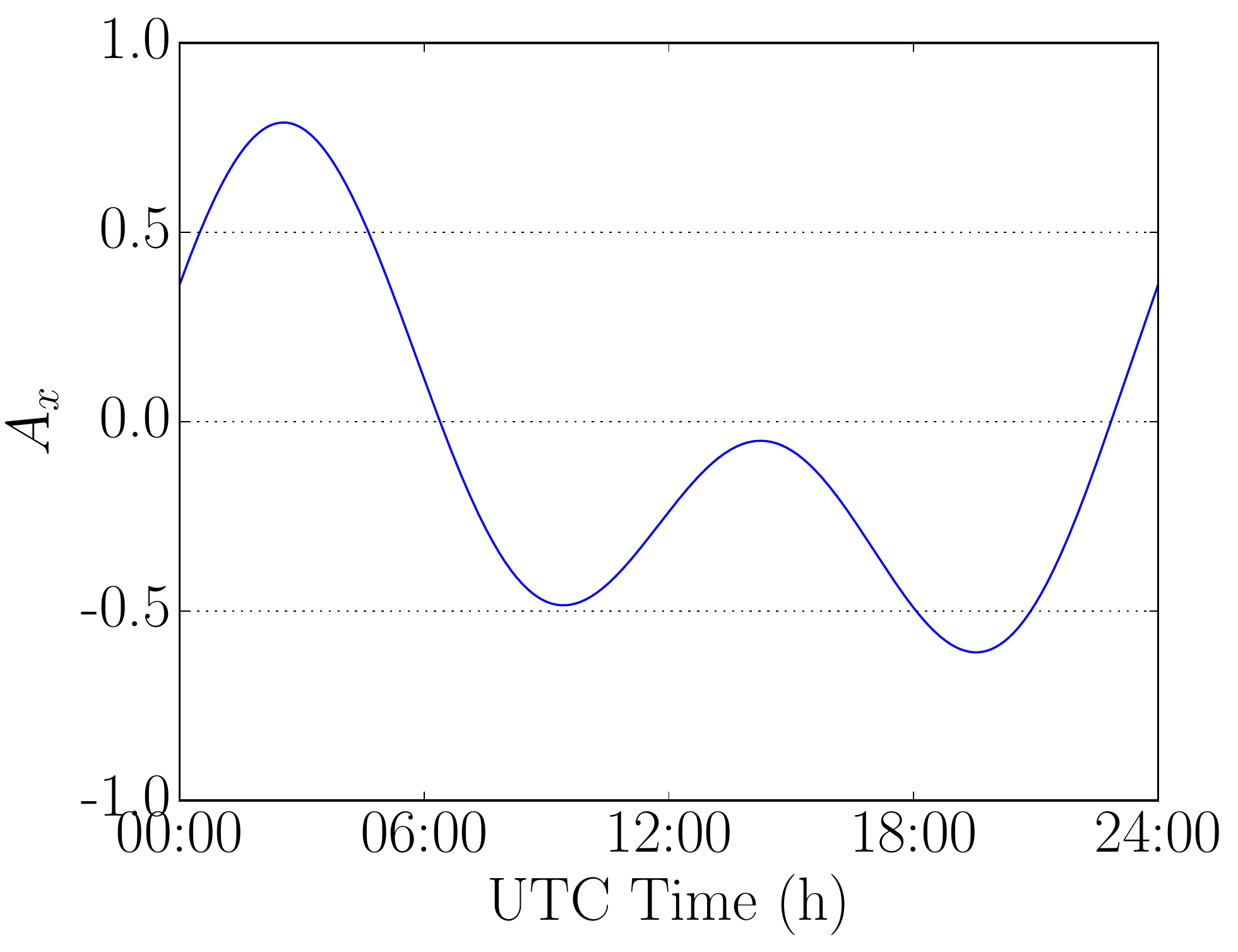}
                \caption{Vector x (x)}
                \label{fig:xz}
        \end{subfigure}%
	\hfill
        \begin{subfigure}[b]{0.33\textwidth}
		\centering
                \includegraphics[width=\textwidth]{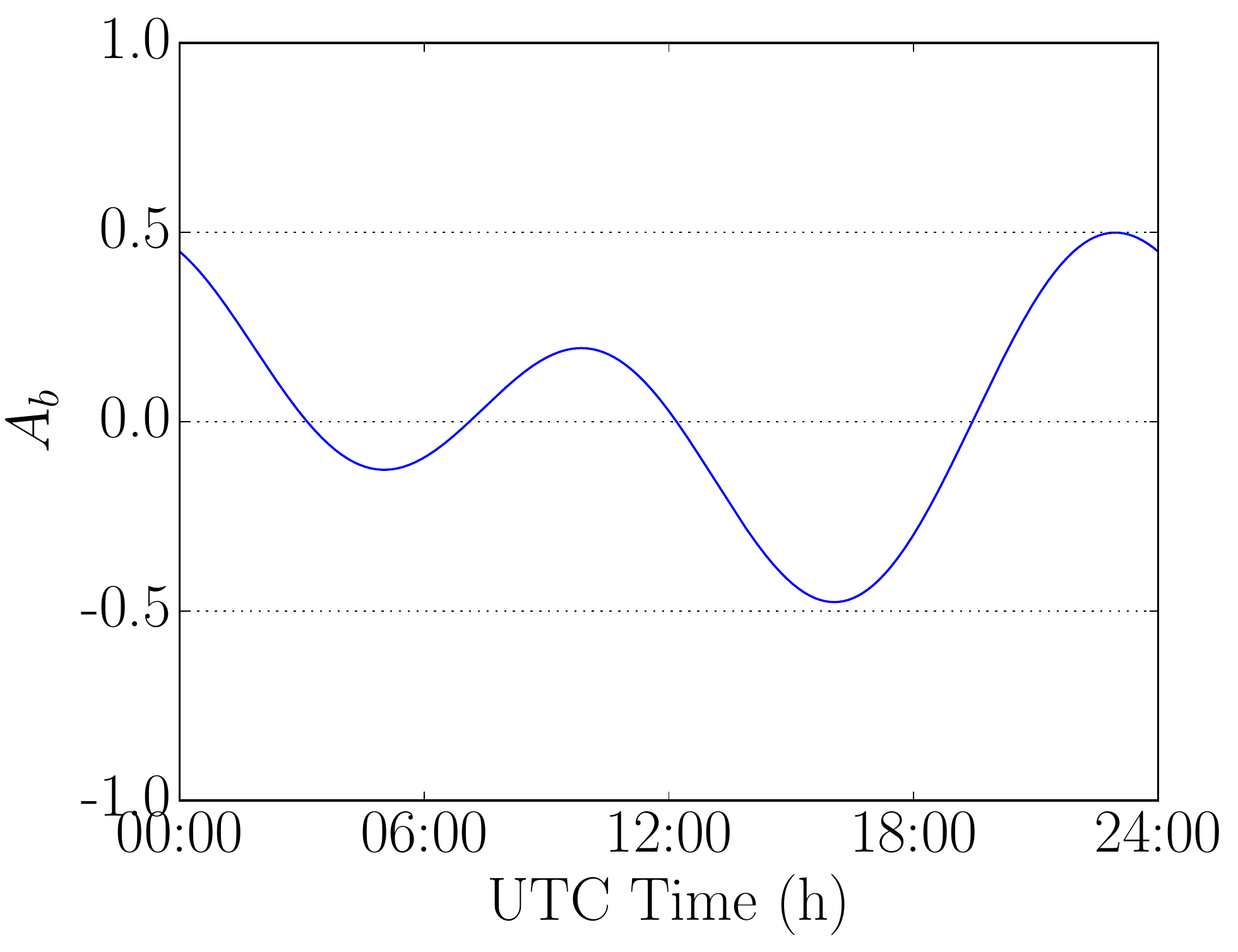}
                \caption{Breathing (b)}
                \label{fig:br}
        \end{subfigure}%
	\\
        \begin{subfigure}[b]{0.33\textwidth}
		\centering
                \includegraphics[width=\textwidth]{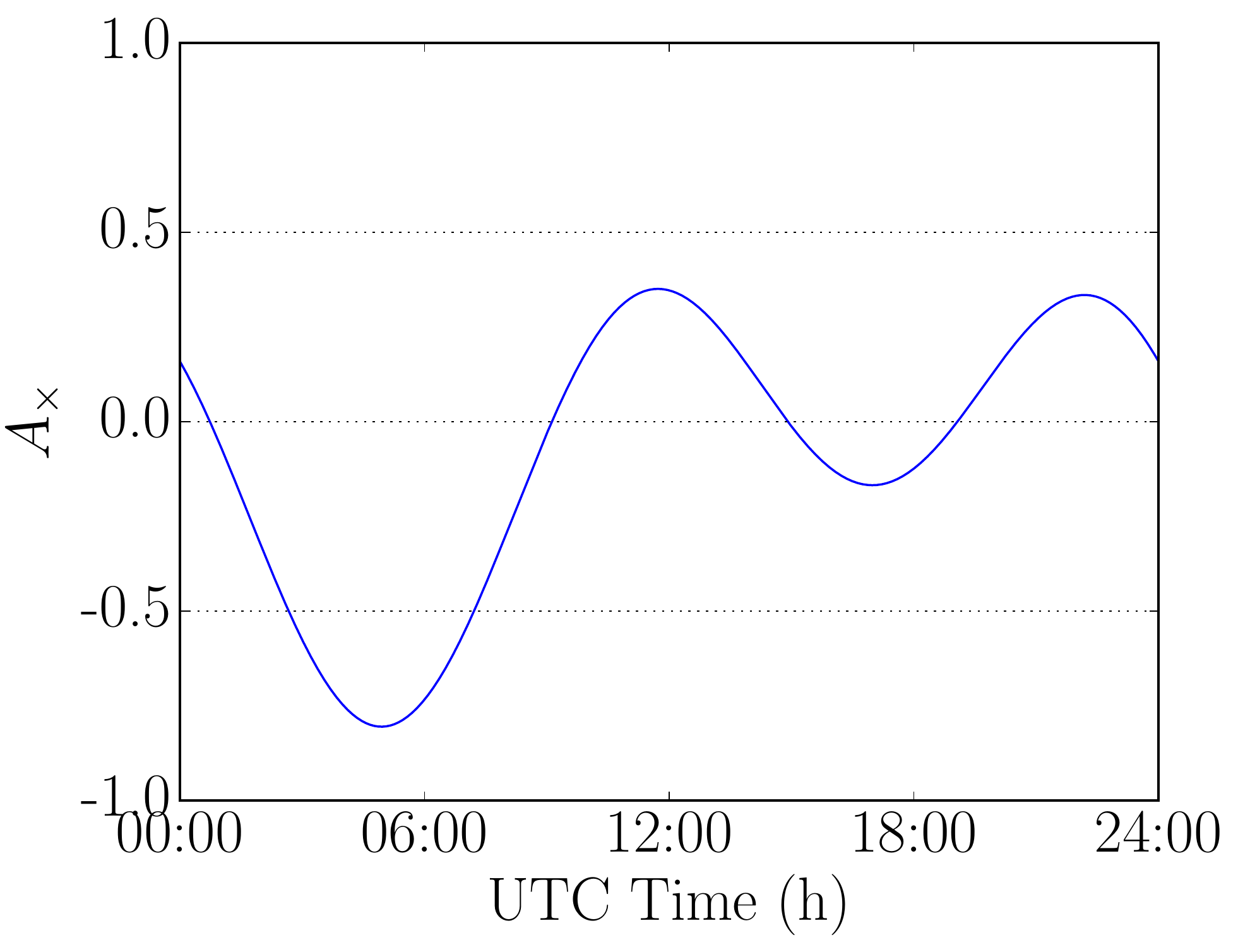}
                \caption{Cross ($\times$)}
                \label{fig:cr}
        \end{subfigure}%
        \hfill
        \begin{subfigure}[b]{0.33\textwidth}
		\centering
                \includegraphics[width=\textwidth]{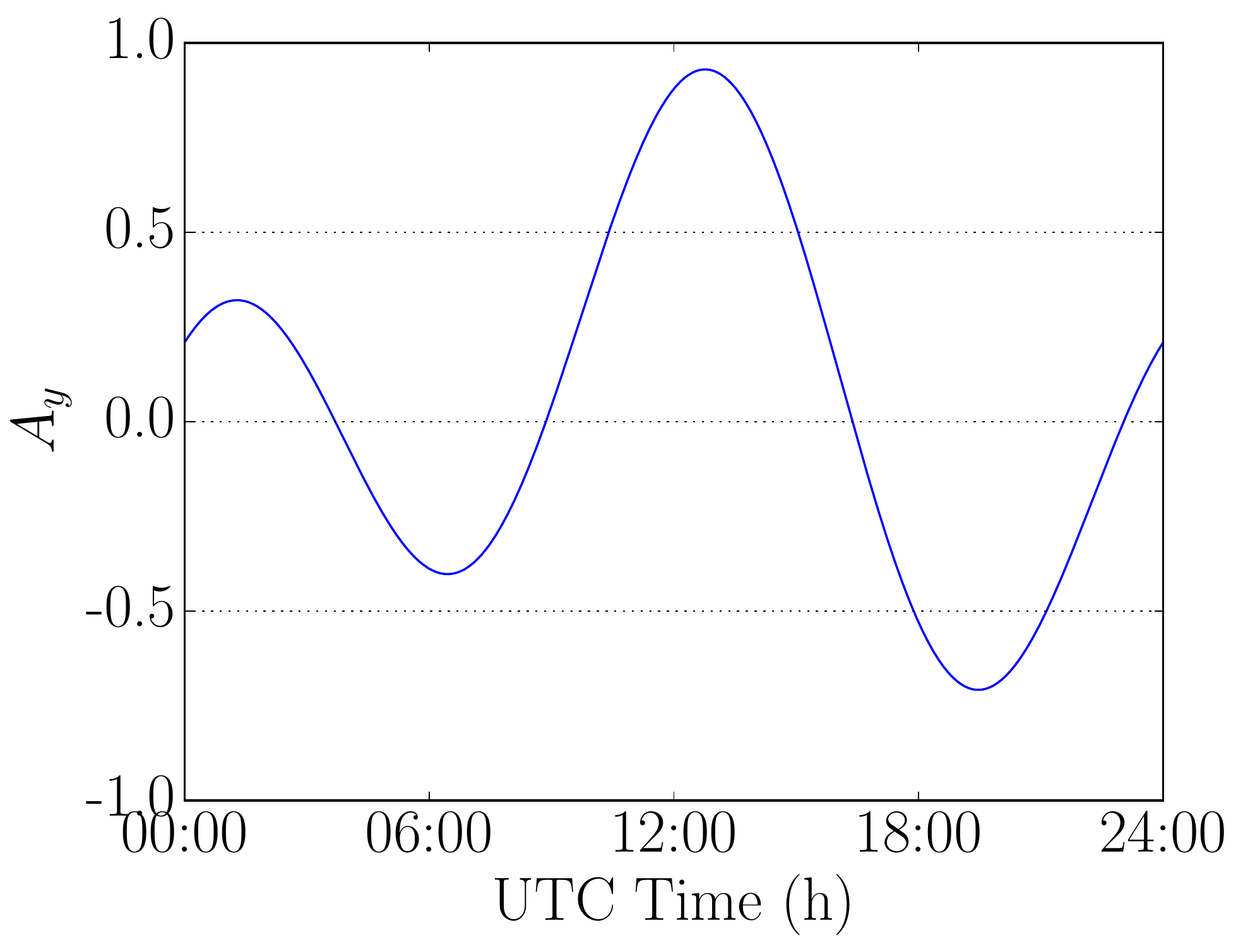}
                \caption{Vector y (y)}
                \label{fig:yz}
        \end{subfigure}%
	\hfill
        \begin{subfigure}[b]{0.329\textwidth}
		\centering
                \includegraphics[width=\textwidth]{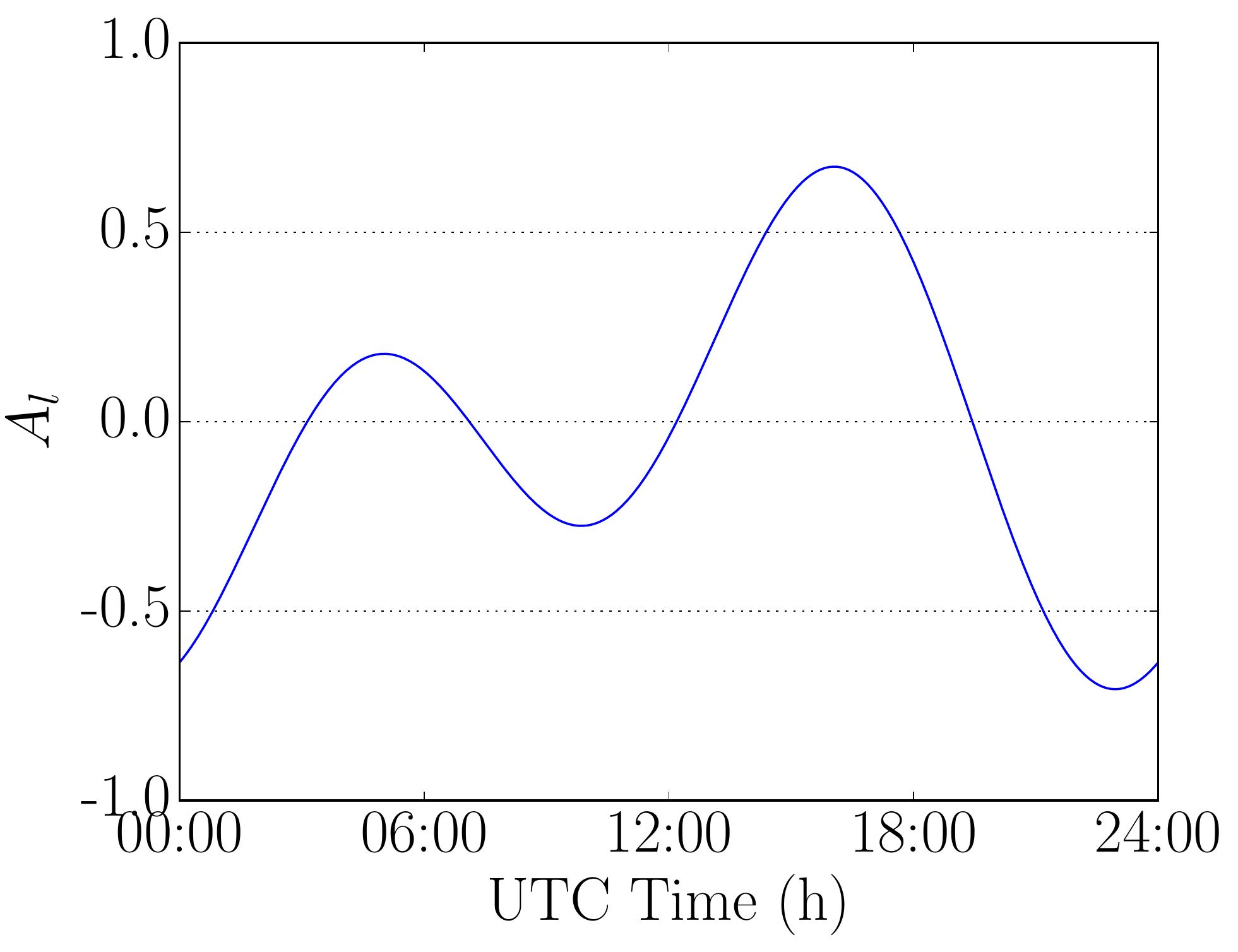}
                \caption{Longitudinal (l)}
                \label{fig:lo}
        \end{subfigure}
        \caption{LHO response $A_p(t)$, eq.~(\ref{eq:response}), to different polarizations from the Crab (PSR J0534+2200), from 00:00 UTC to 24:00 UTC.}\label{fig:polarizations}
\end{figure*}

If a detector is characterized by its unit arm--direction vectors (${\bf d}_x$ and ${\bf d}_y$, with ${\bf d}_z$ the detector zenith), its differential--arm response $A_p$ to a wave of polarization $p$ is:
% In LAL: https://www.lsc-group.phys.uwm.edu/daswg/projects/lal/nightly/docs/html/_create_detector_8c_source.html
\beq \label{eq:response}
A_{p} = \frac{1}{2} \left( {\bf d}_x \otimes {\bf d}_x - {\bf d}_y \otimes {\bf d}_y \right) : {\bf e}^{p},
\eeq
where the colon indicates double contraction. As a result, eqs.~(2-13) imply:
\begin{equation} \label{eq:A+}
A_{+}=\frac{1}{2}\left[({\bf w}_x \cdot {\bf d}_x)^2-({\bf w}_x \cdot {\bf d}_y)^2-({\bf w}_y \cdot {\bf d}_x)^2+({\bf w}_y \cdot {\bf d}_y)^2 \right],
\end{equation}
\begin{equation} \label{eq:AX}
A_{\times}=({\bf w}_x \cdot {\bf d}_x) ({\bf w}_y \cdot {\bf d}_x)-({\bf w}_x \cdot {\bf d}_y) ({\bf w}_y \cdot {\bf d}_y),
\end{equation}
\begin{equation} \label{eq:Ax}
A_{\rm x}= ({\bf w}_x \cdot {\bf d}_x) ({\bf w}_z \cdot {\bf d}_x)- ({\bf w}_x \cdot {\bf d}_y) ({\bf w}_z \cdot {\bf d}_y),
\end{equation}
\begin{equation} \label{eq:Ay}
A_{\rm y}= ({\bf w}_y \cdot {\bf d}_x) ({\bf w}_z \cdot {\bf d}_x)- ({\bf w}_y \cdot {\bf d}_y) ({\bf w}_z \cdot {\bf d}_y),
\end{equation}
\begin{equation} \label{eq:Ab}
A_{\rm b}= \frac{1}{2} \left[ ({\bf w}_x \cdot {\bf d}_x)^2-({\bf w}_x \cdot {\bf d}_y)^2+({\bf w}_y \cdot {\bf d}_x)^2-({\bf w}_y \cdot {\bf d}_y)^2\right],
\end{equation}
\begin{equation} \label{eq:Al}
A_{\rm l}=\frac{1}{\sqrt{2}}\left[ ({\bf w}_z \cdot {\bf d}_x)^2- ({\bf w}_z \cdot {\bf d}_y)^2 \right].
\end{equation}
Accounting for the time dependence of the arm vectors due to the rotation of the Earth, eqs.~(\ref{eq:A+}-\ref{eq:Al}) can be used to compute $A_p(t)$ for any value of $t$. In fig.~\ref{fig:polarizations} we plot these responses for the LIGO Hanford Observatory (LHO) observing the Crab pulsar, over a sidereal day (the pattern repeats itself every day). Note that the b and l patterns are degenerate ($A_{\rm b} = -\sqrt{2} A_{\rm l}$), which means they are indistinguishable up to an overall constant.

Although the antenna patterns are $\psi$--dependent, a change in this angle amounts to a rotation of $A_+$ into $A_\times$ or of $A_x$ into $A_y$, and vice--versa. If the orientation of the source is changed such that the new polarization is $\psi' = \psi + \Delta\psi$, where $\psi$ is the original polarization angle and $\Delta\psi\in \left[0,2\pi\right]$, it is easy to check that the new antenna patterns can be written \cite{Blaut2012}:
\beq \label{eq:A+rot}
A'_+ = A_+ \cos{2\Delta\psi} + A_\times \sin{2\Delta\psi},
\eeq
\beq \label{eq:AXrot}
A'_\times = A_\times \cos{2\Delta\psi} - A_+ \sin{2\Delta\psi},
\eeq
\beq \label{eq:Axrot}
A'_x = A_x \cos{\Delta\psi} + A_y \sin{\Delta\psi},
\eeq
\beq \label{eq:Ayrot}
A'_y = A_y \cos{\Delta\psi} - A_x \sin{\Delta\psi},
\eeq
\beq
A'_b = A_b,
\eeq
\beq \label{eq:Alrot}
A'_l = A_l,
\eeq
and the tensor, vector and scalar nature of each polarization becomes evident from the $\psi$ dependence.

\section{Method} \label{sec:method}
\subsection{Data reduction}

\begin{figure*}  [hbtp!]
        \centering
        \begin{subfigure}[c]{0.49\textwidth}
                \centering
                \includegraphics[width=\textwidth]{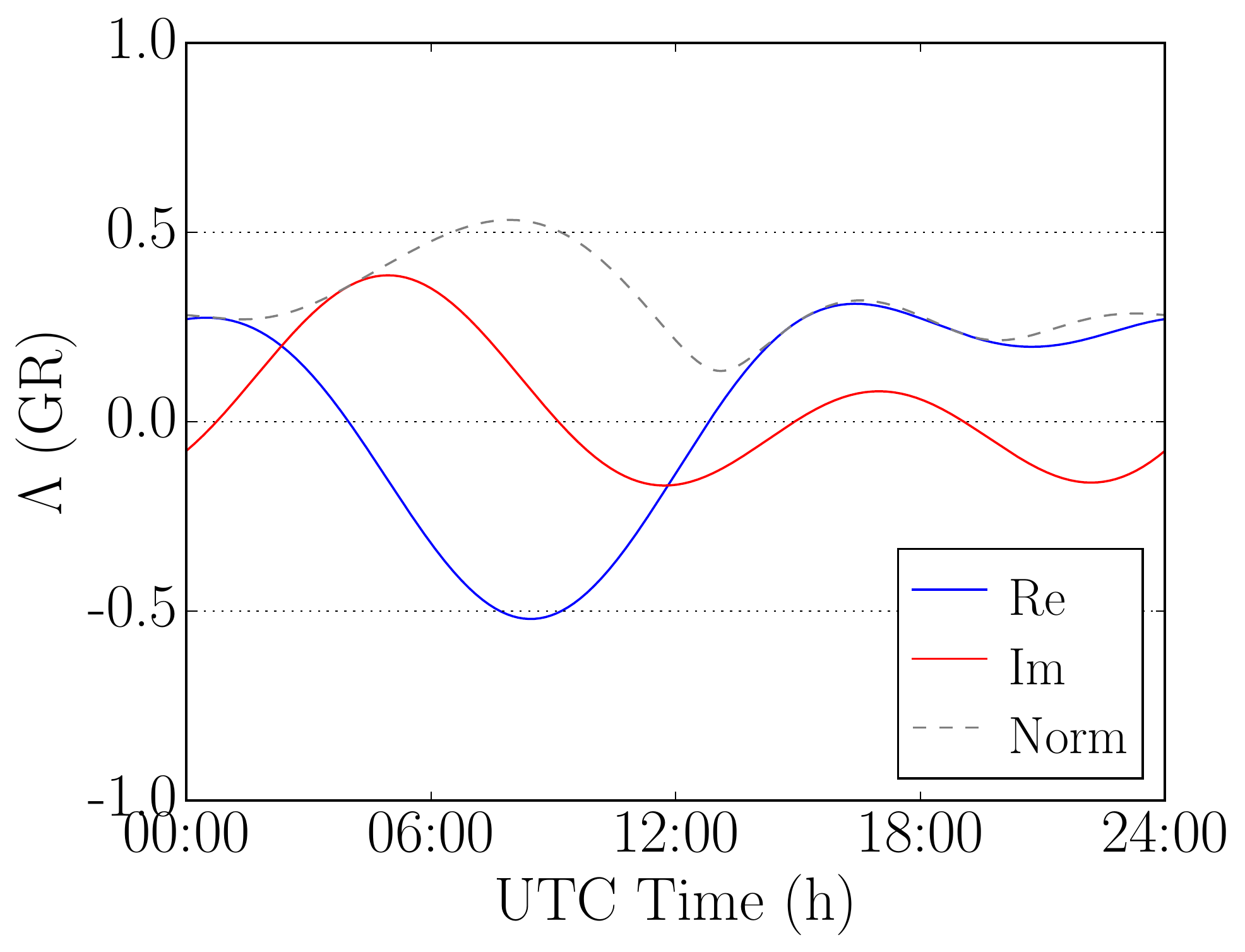}
        \end{subfigure}%
        \hfill
        \begin{subfigure}[c]{0.49\textwidth}
                \centering
                \includegraphics[width=\textwidth]{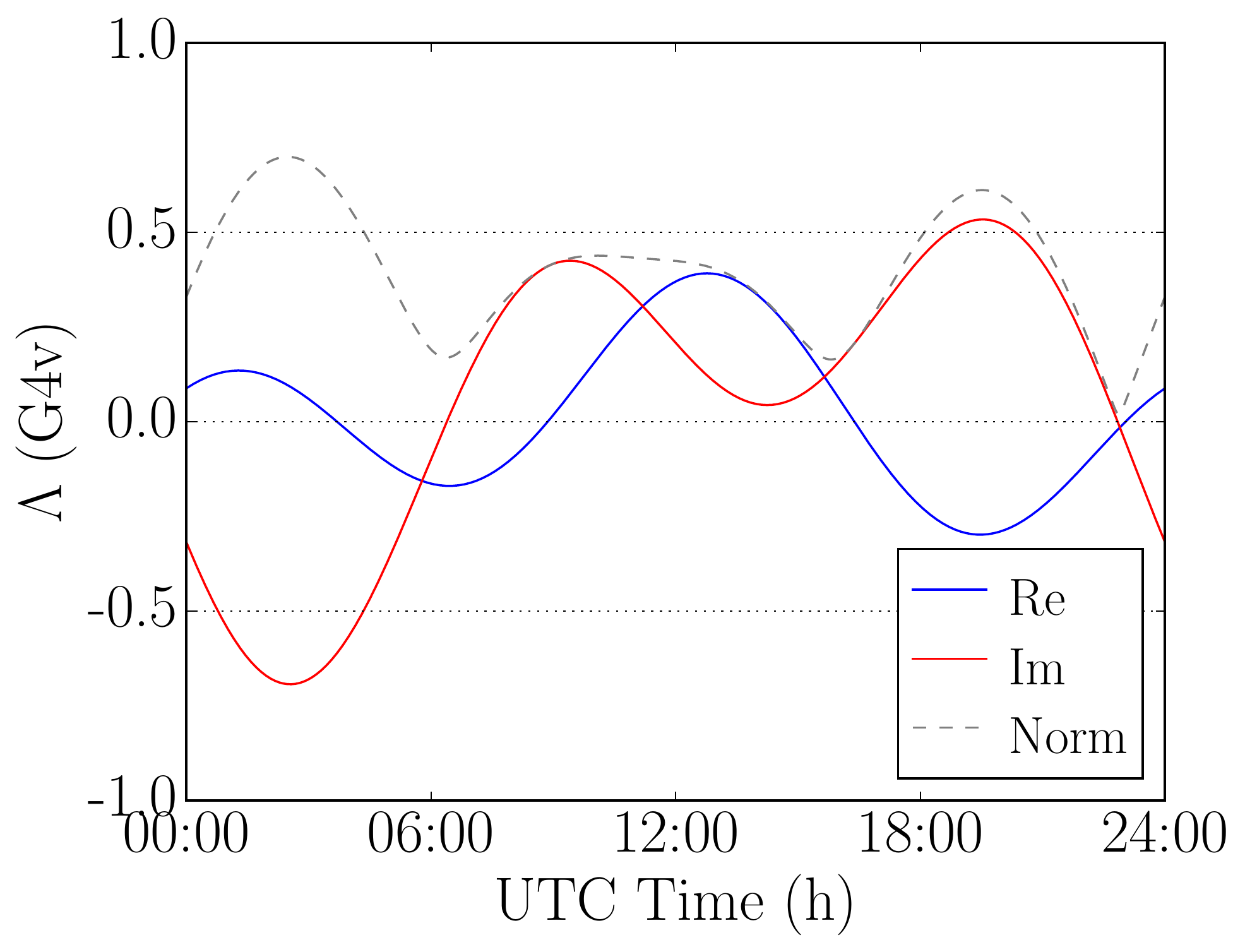}
        \end{subfigure}
        \caption{Simulated GR (left) and G4v (right) heterodyned Crab signals as seen by LHO. The templates are generated from eq.~(\ref{eq:het-signal}) with the model parameters given in eqs.~(\ref{eq:a+}--\ref{eq:ayz}) and setting $h_0=1,~\phi_0=0$. The solid curves represent the real (blue) and imaginary (red) parts, while the dashed curve corresponds to the complex norm.}
\label{fig:signals}
\end{figure*}

For some set of interferometric data, we would like to detect CGW signals from a given source, regardless of their polarization, and to reliably distinguish between the different modes. Because detector response is the only factor distinguishing CGW polarizations, all the relevant information is encoded in the amplitude modulation of the signal. As a result, it suffices to consider a narrow frequency band around the GW frequency and the data can be considerably reduced following the complex heterodyne method developed in \cite{Niebauer1993} and \cite{Dupuis2005}. 

A signal of the form of eq.~(\ref{eq:cgw}) can be re--written as
\beq
h(t) = \Lambda(t) e^{i\phi(t)} + \Lambda^*(t) e^{-i\phi(t)},
\eeq
\beq \label{eq:het-signal}
\Lambda(t) = \frac{1}{2} \sum \limits_{p=1}^5 a_p e^{i\phi_p+i\phi_0}A_p(t) ,
\eeq
with $*$ indicating complex conjugation and $\phi(t)$ as given in eq.~(\ref{eq:phase}). Note that we have slightly simplified the notation in eq.~(\ref{eq:het-signal}) by renaming $\phi^{\rm gw}_0 \rightarrow \phi_0$. Also, the summation is over only five values of \textit{p} because the breathing and longitudinal polarizations are indistinguishable to the detectors.

The key of the heterodyne method is that, since we can assume the phase evolution is well--known from electromagnetic observations (ephemerides obtained through the pulsar timing package TEMPO2 \cite{Edwards2006}), we can multiply our data by $\exp{\left[-i\phi(t)\right]}$ (heterodyning) so that the signal therein becomes
\beq \label{eq:het-data}
h'(t)\equiv h(t) e^{-i\phi(t)} =\Lambda(t) + \Lambda^*(t) e^{-i2\phi(t)}
\eeq
and the frequency modulation of the first term is removed, while that of the second term is doubled. A series of low--pass filters can then be used to remove the quickly--varying term, which enables the down--sampling of the data by averaging over minute--long time bins. As a result, we are left with $\Lambda(t)$ only and eq.~(\ref{eq:het-signal}) becomes the template of our complex--valued signal. One period of such GR and G4v signals coming from the Crab are presented as seen by LHO in fig.~\ref{fig:signals}.

\begin{figure*} [hbtp]
        \centering
                \includegraphics[width=\textwidth]{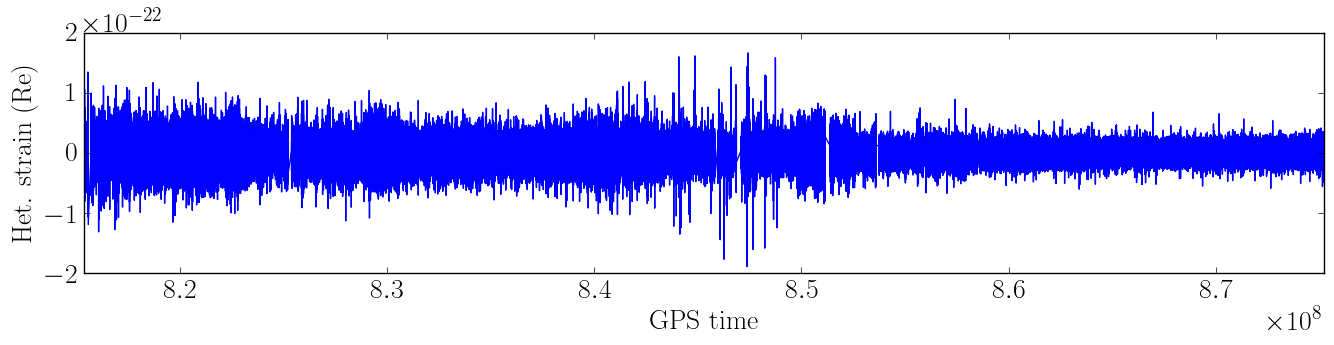}
               \caption{Real part of LIGO Science Run 5 Hanford 4km detector (H1) minute--sampled data prepared for the Crab spanning approximately two years. A signal in these data would be described by eq.~(\ref{eq:data}).}
                \label{fig:finehetS5reH1}
\end{figure*}

From eq.~(\ref{eq:het-data}) we see that, in the presence of a signal, the heterodyned and down-sampled noisy detector strain data $B_{k}$ for the $k^{th}$ minute-long time bin (which can be labeled by GPS time of arrival) are expected to be of the form:
\begin{equation} \label{eq:data}
B_{\rm expected}(t_{k})= \frac{1}{2} \sum\limits_{p=1}^{5} a_p(t_k)  e^{i\phi_p+i\phi_0} A_p(t_k)+n(t_{k}),
\end{equation}
where $n(t_{k})$ is the heterodyned, averaged complex noise in bin \textit{k}, which carries no information about the GW signal. As an example, fig.~\ref{fig:finehetS5reH1} presents the real part of actual data heterodyned and filtered for the Crab pulsar. We can clearly see already that the data are non--stationary, an issue addressed in the section \ref{sec:search} and appendix \ref{sec:stats}.

\subsection{Search}  \label{sec:search}

Given data in this form, we analyze it to obtain the parameters of a signal that would best fit the data and then incorporate the results into the frequentist analysis described in section \ref{sec:analysis}. Regressions are performed by minimizing the $\chi^2$ of the system (same as a matched--filter). For certain template $T(t_k)$, this is:
\begin{equation}
\chi ^2=\displaystyle\sum\limits_{k=0}^{N}\ \left[T(t_k)-B(t_{k})\right]^2 /{\sigma_k ^2},
\label{eq:chi}
\end{equation}
where $\sigma_k$ is the estimate standard deviation of the noise in the data at time $t_k$. In the presence of Gaussian noise, the $\chi^2$ minimization is equivalent to a maximum likelihood analysis.

Any linear template $T$ can be written as a linear combination of certain basis functions $f_i$, so that $T(t) =\displaystyle\sum\limits_{i} \tilde{a}_i f_i(t)$ and each $\tilde{a}_i$ is found as a result of minimizing (\ref{eq:chi}). For instance, $T(t_k)$ could be constructed in the from of eq.~(\ref{eq:het-signal}). In such model--dependent searches, the antenna patterns are the basis set, i.e.~$\{f_i\}=\{A_p\}$, and the $\tilde{a}_i$ weights correspond to the $a_p \exp{\left(i\phi_p\right)}$ prefactors. (From here on, the tilde denotes the coefficient that is fitted for, rather than its predicted value.)

The regression returns a vector ${\bf \tilde{a} }$ containing the values of the $\tilde{a}_i$'s that minimize eq.~(\ref{eq:chi}). These quantities are complex--valued and encode the relative amplitude and phase of each contributing basis. From their magnitude, we define the overall \emph{recovered signal strength} to be:
\beq \label{eq:hrec}
\hrec = | {\bf \tilde{a} }|.
\eeq
The significance of the fit is evaluated through the covariance matrix $C$. This can be computed by taking the inverse of $A^T A$, where $A$ is the design matrix of the system (built from the $f_i$ set). In particular, we define the \emph{significance} of the resulting fit (signal SNR) as
\begin{equation} \label{eq:sig}
s = \sqrt{ \mathbf{\tilde{a}}^{\dagger} C^{-1}  \mathbf{\tilde{a}} },
\end{equation}
where $\dagger$ indicates Hermitian conjugation.

$\chi^2$--minimizations have optimal performances when the noise is Gaussian. However, although the central limit theorem implies that the averaged noise in (\ref{eq:data}) should be normally distributed, actual data is far from this ideal (see fig.~\ref{fig:finehetS5reH1}). In fact, the quality of the data changes over time, as it is contingent on various instrumental factors. The time series is plagued with gaps and is highly non--stationary. This makes estimating $\sigma_k$ non--trivial.

\begin{figure} [hbtp]
\centering
\includegraphics[scale=.5]{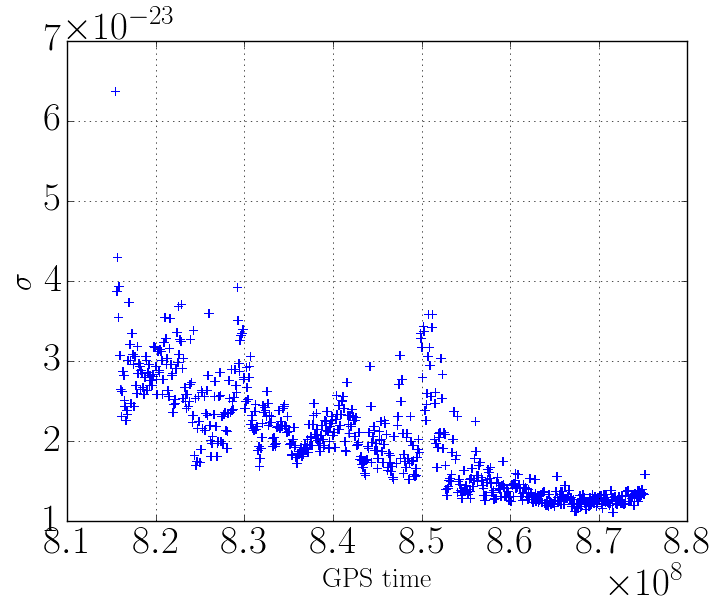}
\caption{Daily standard deviation of S5 H1 data heterodyned for the  Crab pulsar (fig.~\ref{fig:finehetS5reH1}).}\label{fig:std}
\end{figure}

As done in regular CW searches \cite{LSC2010}, we address this problem by computing the standard deviation for the data corresponding to each sidereal day throughout the data run, rather than for the series as a whole. This method improves the analysis because the data remains relatively stable over the course of a single day, but not throughout longer periods of time (see appendix \ref{sec:stats}). Furthermore, noisier days have less impact on the fit, because $\sigma_k$ in eq.~(\ref{eq:chi}) will be larger. The evolution of the daily value of the standard deviation for H1 data heterodyned for the Crab pulsar is presented in fig.~\ref{fig:std}.

\subsubsection{Model--dependent}

In a model--dependent search, a particular physical model is assumed in order to create a template based on eq.~(\ref{eq:het-signal}). In the case of GR, if $\psi$ and $\iota$ are known, it is possible to construct a template with only one complex--valued free parameter $\tilde{h}_0$:
\begin{align} \label{eq:tempGRlocked}
T_{\rm GR}(t) = \tilde{h}_0 \frac{1}{2}&\left[\frac{1}{2}(1+\cos^2\iota) A_+(t; \psi) +\right. \nonumber \\
&+\left.\cos\iota A_\times (t; \psi) e^{-i \pi/2}\right],
\end{align}
where the factor of 2 comes from the heterodyne, cf.~eq.~(\ref{eq:het-signal}). Similarly for G4v:
\begin{align} \label{eq:tempG4vlocked}
T_{\rm G4v}(t) = \tilde{h}_0 \frac{1}{2}\left[\sin\iota~e^{-i \pi/2}A_{\rm x}(t; \psi) + \sin\iota\cos\iota A_{\rm y} (t; \psi) \right],
\end{align}
Analogous templates could be constructed for scalar–-tensor theories, or any other model. In the former case, there would be a second free parameter to represent the unknown scalar contribution.

However, as mentioned in section \ref{sec:background}, even in the case of the best studied pulsars we know $\iota$ only in absolute value. This ambiguity creates the need to use two model--dependent templates like eqs.~(\ref{eq:tempGRlocked}, \ref{eq:tempG4vlocked}): one corresponding to $\iota$ and one to $\pi-\iota$. Note that the indeterminacy of $\psi$ is absorbed by the overall phase of $\tilde{h}_0$, so it has no effect on the template. Thus, if the ambiguity in $\iota$ is accounted for, the overall signal strength $h_0$ and the angle $\phi_0$ can be inferred directly from the angle and phase of $\hrec = \tilde{h}_0$.

In most cases, $\psi$ and $\iota$ are completely unknown. It is then convenient to regress to each antenna pattern independently, allowing for two free parameters. This can be done by computing the antenna patterns assuming any arbitrary value of the polarization angle, say $\psi=0$. Indeed, eqs.~(\ref{eq:A+rot}--\ref{eq:Alrot}) guarantee that the subspace of tensor, vector or scalar antenna patterns for \emph{all} $\psi$ is spanned by a pair of corresponding tensor, vector or scalar antenna patterns assuming any \emph{particular} $\psi$.

In the case of GR, this means we can use a template
\beq \label{eq:tempGR}
T_{\rm GR}(t)=\tilde{\alpha}_+A_+(t;\psi=0)/2 + \tilde{\alpha}_\times A_\times(t;\psi=0)/2
\eeq
with two complex weights $\tilde{\alpha}$'s to be determined by the minimization. In the presence of a signal and in the absence of noise, eqs.~(\ref{eq:A+rot}, \ref{eq:AXrot}) indicate that the values returned by the fit would be a function of the \emph{actual}, unknown $\psi$ and $\iota$:
\beq
\alpha_+ = a_+(\iota) e^{i\phi_0} \cos{2\psi} - a_\times(\iota) e^{i\phi_0-i\pi/2} \sin{2\psi},
\eeq
\beq
\alpha_\times = a_\times(\iota) e^{i\phi_0-i\pi/2} \cos{2\psi} + a_+(\iota) e^{i\phi_0}\sin{2\psi},
\eeq
with the $\alpha(\iota)$'s as given in eqs.~(\ref{eq:a+}, \ref{eq:ax}).

Again, a (semi--) model--dependent template, like eq.~(\ref{eq:tempGR}), can be constructed for any given theory by selecting the corresponding antenna patterns to be used as basis for the regression. For G4v, this would be:
\beq \label{eq:tempG4v}
T_{\rm G4v}(t)=\tilde{\alpha}_xA_x(t;\psi=0)/2 + \tilde{\alpha}_y A_y(t;\psi=0)/2
\eeq
with two complex weights $\tilde{\alpha}$'s to be determined by the minimization. As before, in the presence of a signal and in the absence of noise, eqs.~(\ref{eq:Axrot}, \ref{eq:Ayrot}) indicate that the values returned by the fit would be a function of the \emph{actual}, unknown $\psi$ and $\iota$:
\beq
\alpha_x = a_x(\iota) e^{i\phi_0-i\pi/2} \cos{\psi} - a_y(\iota) e^{i\phi_0} \sin{\psi},
\eeq
\beq
\alpha_y = a_y(\iota) e^{i\phi_0} \cos{\psi} + a_x(\iota) e^{i\phi_0-i\pi/2}\sin{\psi}.
\eeq

In this case, we cannot directly relate our recovered strength to $h_0$ and the framework does not allow to carry out parameter estimation. The proper way to do that is using Bayesian statistics, marginalizing over the orientation parameters. Since we are mostly interested in quantifying our ability to detect alternative signals rather than estimating source parameters, we do not cover such methods here. However, it would be straightforward to incorporate our generalized likelihoods (as given by our templates) into a full Bayesian analysis (cf.~\cite{Dupuis2005}).

\subsubsection{Model--independent}

In a model--independent search, the regression is performed using all five non--degenerate antenna patterns and the phases between the $A_p$'s are not constrained. Thus,
\beq
T_{\rm indep}(t) = \displaystyle\sum\limits_{p=1}^{5}\tilde{a}_{p}A_{p}(t).
\eeq
Because we do not consider any particular model, there is no information about the relative strength of each polarization; hence, the $\tilde{a}_p$'s are unconstrained. Again, eqs.~(\ref{eq:A+rot}--\ref{eq:Alrot}) enable us to compute the antenna patterns for any value of $\psi$.

By calculating the necessary inner products, it can be shown that a regression to the \emph{antenna pattern basis},
\beq
\left\{A_+, ~A_\times,~ A_{\rm x}, ~A_{\rm y}, ~A_{\rm b}\right\},
\eeq
is equivalent to a regression to the \emph{sidereal basis},
\begin{equation} \label{eq:sidbasis}
\left\{1,~\cos{\omega t},~\cos{2\omega t},~\sin{\omega t},~\sin{2\omega t} \right\},
\end{equation}
where $\omega = 2\pi / (86 164 ~{\rm s})$ is the sidereal rotational frequency of the Earth. This is an orthogonal basis which spans the space of the antenna patterns. In this basis,
\beq \label{eq:tempSid}
T_{\rm indep}(t) = \displaystyle\sum\limits_{i=1}^{5}\tilde{a}_{i}f_{i}(t).
\eeq
with $f_i$ representing the set in (\ref{eq:sidbasis}). This is the same basis set used in so--called \emph{5-vector searches} \cite{Astone2010}.

Because they span the same space, using either basis set yields the same results with the exact same significance, as defined in eq.~(\ref{eq:sig}). Furthermore, the weights obtained as results of the fit can be converted back and forth between the two bases by means of a time--independent coordinate transformation matrix.

A model--independent search is sensitive to all polarizations, but is prone to error due to noise when distinguishing between them. It also has more degrees of freedom (compared with a pure-GR template) that can respond to noise fluctuations, resulting in a search that is less sensitive to pure-GR signals. However, the analysis can be followed by model--dependent searches to clarify which theory fits with most significance.

\begin{figure*} [!hbtp]
        \centering
        \begin{subfigure}[c]{0.33\textwidth}
                \centering
                \includegraphics[width=\textwidth]{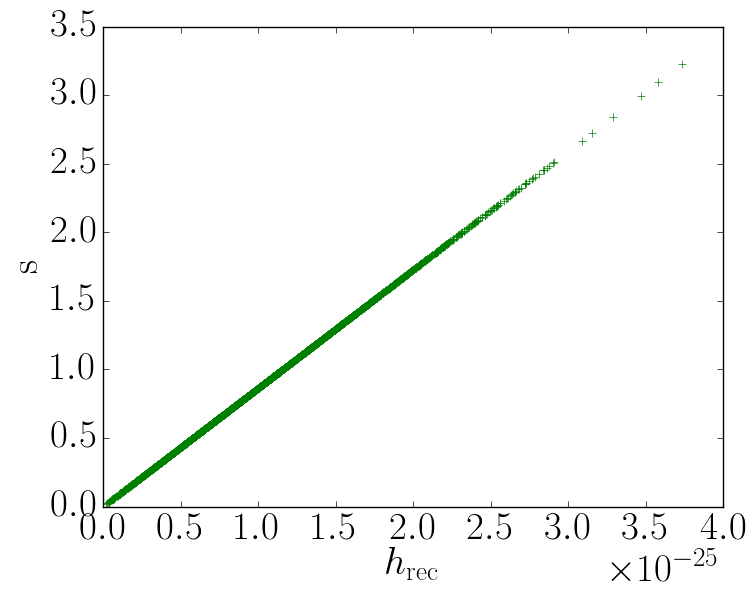}
        \end{subfigure}%
	\hfill
        \begin{subfigure}[c]{0.33\textwidth}
                \centering
                \includegraphics[width=\textwidth]{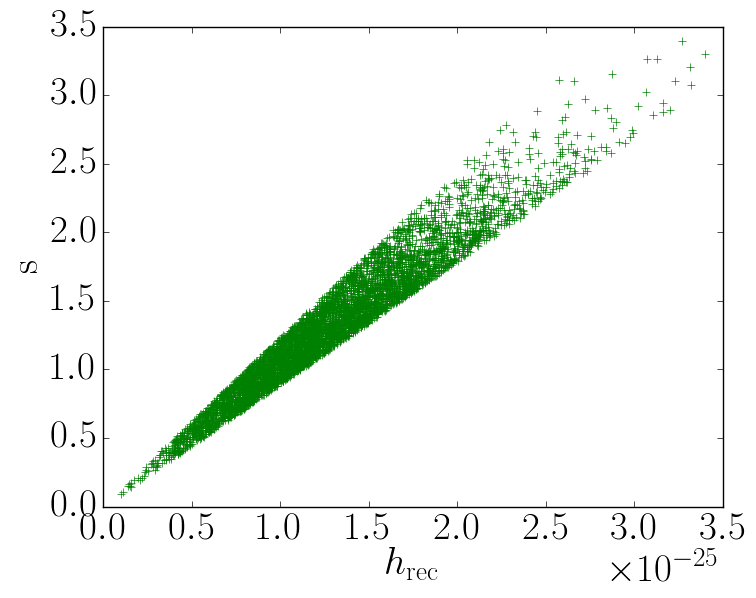}
        \end{subfigure}%
	\hfill
        \begin{subfigure}[c]{0.33\textwidth}
                \centering
                \includegraphics[width=\textwidth]{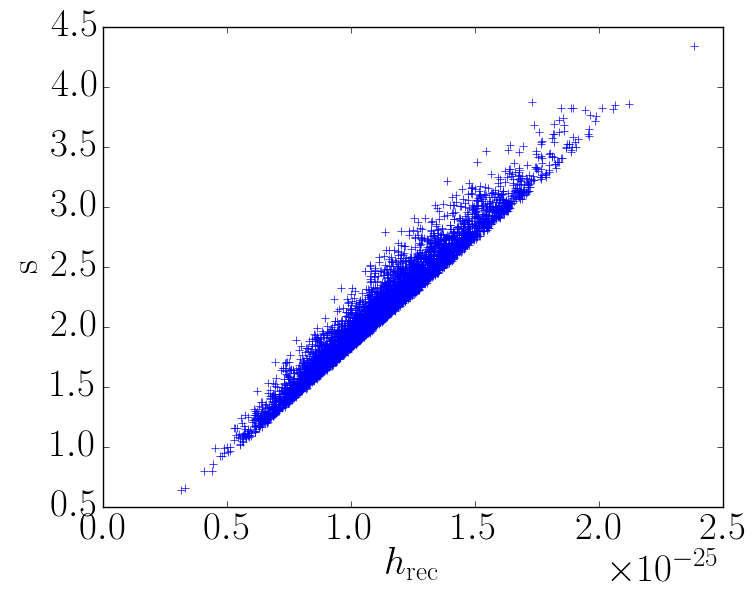}
        \end{subfigure}
        \caption{Significance, eq.~(\ref{eq:sig}), vs.~recovered strength, eq.~(\ref{eq:hrec}), for searches over 5000 noise--only H1 S5 Crab instantiations using model--dependent eq.~(\ref{eq:tempGRlocked}) (left), semi--dependent eq.~(\ref{eq:tempGR})  (center), and independent eq.~(\ref{eq:tempSid})  (right) templates. The model--dependent case assumes \emph{fully} known $\iota$ and $\psi$. Note that the number of degrees of freedom in the regression is manifested in the spread, which is due to noise: templates with a single degree of freedom are less susceptible to noise and the spread is minimal. The two plots on the left were generated using a GR template, but similar results are obtained for G4v.}
\label{fig:hs}
\end{figure*}

\begin{figure} [!hbtp]
\centering
\includegraphics[width=\columnwidth]{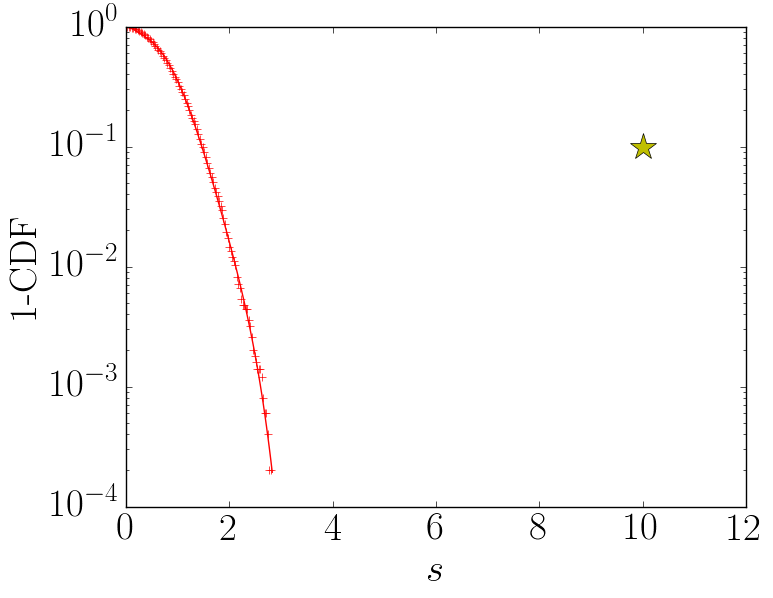}
\caption{Example plot of $p=1-{\rm CDF}$ vs.~the recovery significance for a particular template. A loud injection in noise is manifested as an outlier (star) over the noise--only background (red). Note that the injection is plotted arbitrarily at $p=10^{-1}$.}
\label{fig:pSidH1S5}
\end{figure}

\begin{figure*}
        \centering
        \begin{subfigure}[b]{0.49\textwidth}
                \centering
                \includegraphics[width=\textwidth]{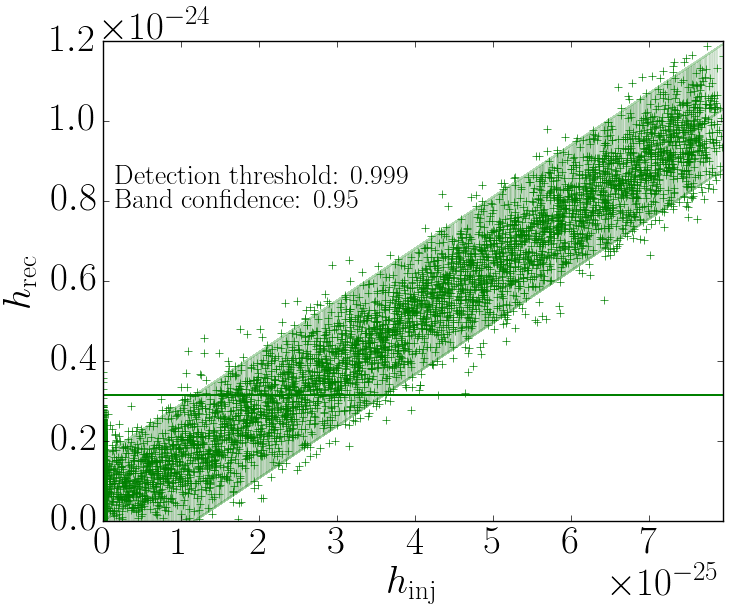}
        \end{subfigure}\hfill
        \begin{subfigure}[b]{0.487\textwidth}
                \centering
                \includegraphics[width=\textwidth]{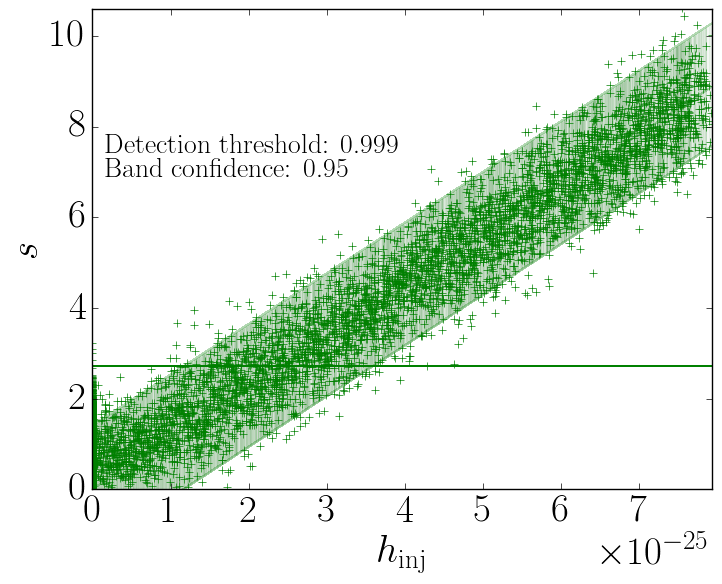}
        \end{subfigure}
        \caption{Neyman plot of recovered signal strength $\hrec$ (left) and significance $s$ (right) vs.~injected strength $\hinj$. In this case, GR signals are recovered with GR templates, but results are qualitatively the same with G4v injections recovered with G4v templates, or either kind of injection recovered with model--independent templates. The collection of points at $\hinj=0$ are noise--only and the detection threshold (horizontal line) is placed above $\fracdetthrsh=99.9\%$ of them. The shaded band includes $\fracbandconf=95\%$ of the data points above the threshold and it is centered on their best--fit line. The fit forced null \emph{y}--intersect.}
\label{fig:injsrch_ex}
\end{figure*}

\section{Analysis} \label{sec:analysis}

We wish to detect any CGW signal originating in a given pulsar, regardless of its polarization in a model--independent way. We can then determine whether the measured polarization content agrees with theoretical predictions. This information can be used to obtain frequentist confidence levels for a potential detection and to generate upper limits for the strength of signals of any polarization potentially buried in the data. 

In order to test the statistical properties of the noisy data filtered through our templates, we produce numerous instantiations of detector noise by taking actual data processed as outlined in section \ref{sec:method} and re--heterodyning over a small band close to the frequency of the original heteredoyne. Any true signal in the data stream is scrambled in the process and what remains is a good estimate of the noise. This allows us to perform searches under realistic conditions with or without injections of simulated signals, while remaining blind to the presence of a true signal. 

By heterodyning at different frequencies, we are able to generate a large number of instantiations of the data. Because our S5 datasets span roughly 1.9 years and are sampled once per minute, our bandwidth is $8.3 \times10^{-3}$ Hz with a lowest resolvable frequency of $1.7 \times 10^{-8}$ Hz. This means we could theoretically re--heterodyne our data at a maximum of $8.3\times10^{-3} / 1.7\times 10^{-8} \approx 4.9\times10^5$ independent frequencies. In our study, we picked $10^4$ frequencies in the $10^{-7}-10^{-3}$ Hz range, avoiding the expected signal frequency of $\sim 10^{-5}$ Hz (period of a sidereal day) and its multiples.

We quantify the results of a particular search by looking at the obtained recovered signal strength, eq.~(\ref{eq:hrec}), and significance, eq.~(\ref{eq:sig}). As expected, these two parameters are strongly correlated (fig.~\ref{fig:hs}). However, the significance is, in the presence of Gaussian noise, a direct indicator of goodness--of--fit and can be used to compare results from templates with different numbers of degrees of freedom.

By performing searches on multiple instantiations of noise--only data, we construct cumulative distribution function (CDF) probability plots showing the distribution of recovered signal strength, eq.~(\ref{eq:hrec}), and significance, eq.~(\ref{eq:sig}), corresponding to a given template. Such plots give the probability that the outcome of the regression is consistent with noise (i.e.~provide $p$--values). As shown in fig.~\ref{fig:pSidH1S5}, an instantiation that contains a loud injected signal becomes manifest in this plot as an outlier. This sort of plot can also be used when searching for an actual signal in the data---namely, when looking at the original, non--reheterodyned series. In that case, the $1-{\rm CDF}$ curve can be extrapolated or interpolated to find the $p$--value corresponding to the significance with which the injection was recovered.

After injecting and retrieving increasingly loud signals with a given polarization content in different background instantiations, we produce plots of recovered strength vs.~injected strength ($\hrec$ vs.~$\hinj$) and significance vs.~injected strength ($s$ vs.~$\hinj$). Recall that injections are of the form of eqs.~(\ref{eq:tempGRlocked}, \ref{eq:tempG4vlocked}). Examples of such plots are presented in fig.~\ref{fig:injsrch_ex}. These plots, and corresponding fits, can be used to assess the sensitivity of a template to certain type of signal, define thresholds for detection and produce confidence bands for recovered parameters. (In the frequentist literature, these plots are sometimes referred to as \emph{Neyman constructions} \cite{Olive2012}.)

We define a horizontal \emph{detection threshold} line above an arbitrary fraction $\fracdetthrsh$ (e.g.,~$\fracdetthrsh= 99.9\%$) of noise--only points (i.e.~points with $\hinj=0$, but  $\hrec\neq0$), so that data points above this line can be considered detected with a $p$--value of $p=1-\fracdetthrsh$ (e.g.,~$p=0.1\%$). For a particular template, this fractional threshold can be directly translated into a significance value $\sigdetthrsh{\fracdetthrsh}$ (e.g.,~$\sigdetthrsh{99.9\%}=2.5$). The sensitivity of the template is related to the number of injections recovered with a significance higher than $\sigdetthrsh{\fracdetthrsh}$. Therefore, for a given $\fracdetthrsh$, a lower  $\sigdetthrsh{\fracdetthrsh}$ means higher sensitivity to true signals.

\newcommand{\fracdetconf}{\alpha_{\rm up}}
For the results of each template, the fractional threshold $\fracdetthrsh$ can also be associated to a strain value. We define this to be the loudness of the minimum injection detected above this threshold with some arbitrary \emph{upper--limit confidence} $\fracdetconf$. This value can be determined from the $s$ vs.~$\hinj$ plot by placing a line parallel to the best fit but to the right of a fraction $\fracdetconf$ of all data points satisfying $0<\hinj$. The intersection of this line with the $\fracdetthrsh$ line occurs at $\hinj=h_{\rm min}^{\fracdetconf}$, which is the strain value above which we can have $\fracdetconf$ confidence that a signal will be detected (i.e.~recovered with significance $s>\sigdetthrsh{\fracdetthrsh}$).

\begin{figure*} [!hbtp]
        \centering
        \begin{subfigure}[b]{0.5\textwidth}
                \centering
                \includegraphics[width=\textwidth]{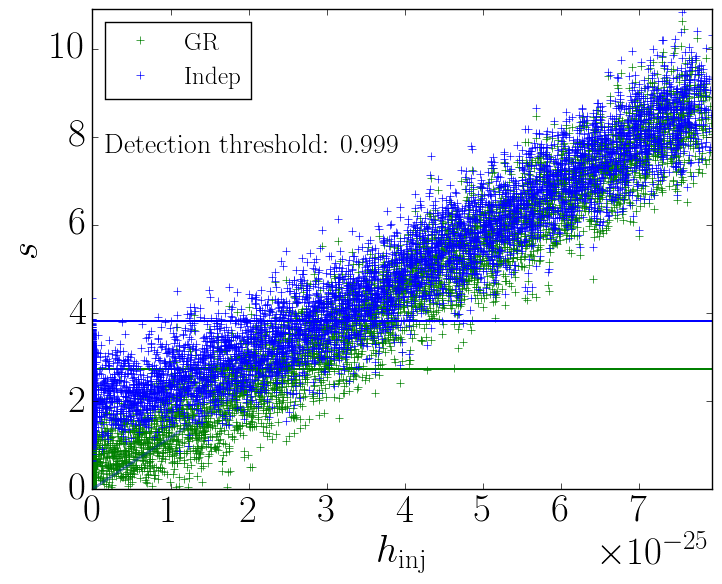}
                \caption{GR injections recovered with GR template (green), eq.~(\ref{eq:tempGR}), and model independent (blue), eq.~(\ref{eq:tempSid}).}
                \label{fig:crabS5gr_indep}
        \end{subfigure}%
        \hfill
        \begin{subfigure}[b]{0.5\textwidth}
                \centering
                \includegraphics[width=\textwidth]{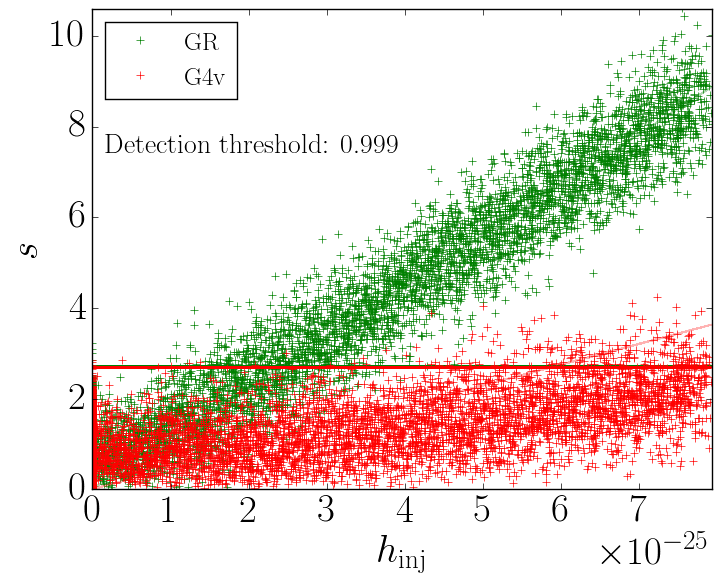}
                \caption{GR injections recovered with GR template (green), eq.~(\ref{eq:tempGR}), and G4v template (red), eq.~(\ref{eq:tempG4v}).}
                \label{fig:crabS5gr_dep}
        \end{subfigure}
\\
        \begin{subfigure}[b]{0.5\textwidth}
                \centering
                \includegraphics[width=\textwidth]{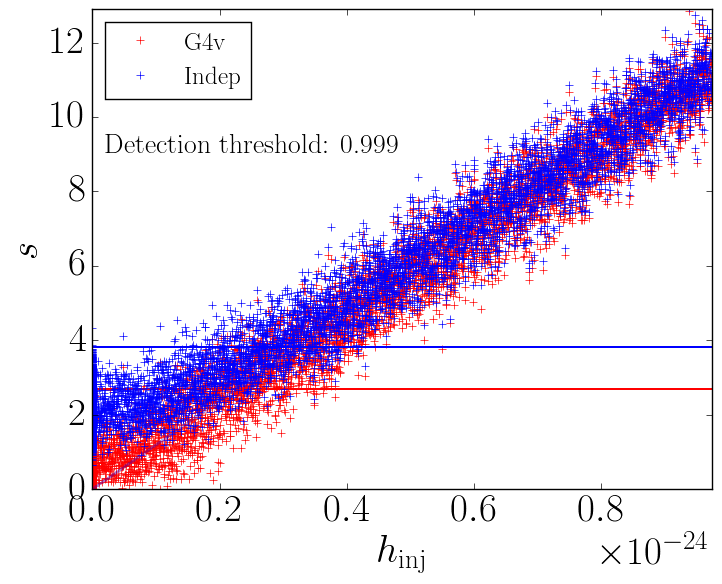}
                \caption{G4v injections recovered with G4v template (red),\\eq.~(\ref{eq:tempG4v}), and model independent (blue), eq.~(\ref{eq:tempSid}).}
                \label{fig:crabS5g4v_indep}
        \end{subfigure}%
        \hfill%
        \begin{subfigure}[b]{0.5\textwidth}
                \centering
                \includegraphics[width=\textwidth]{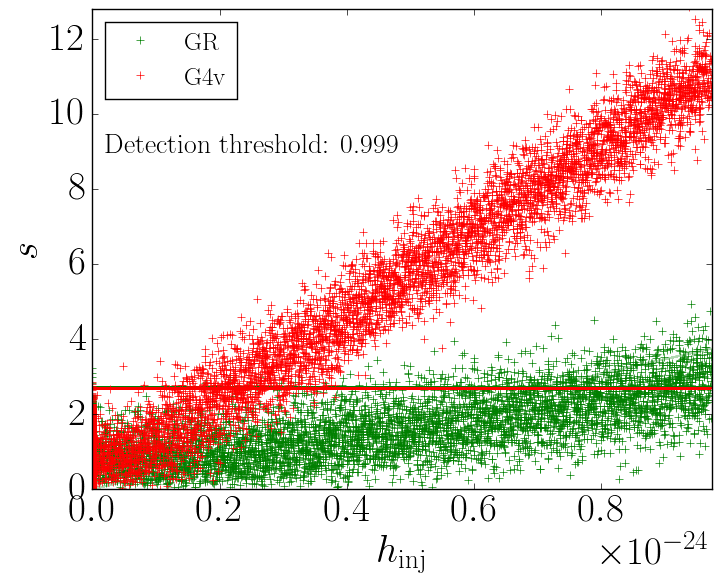}
                \caption{G4v injections recovered with G4v template (red), eq.~(\ref{eq:tempG4v}), and GR template (green), eq.~(\ref{eq:tempGR}).}
                \label{fig:crabS5g4v_dep}
        \end{subfigure}
        \caption{GR (top) and G4v (bottom) injection results of search over LIGO S5 H1 data heterodyned for the Crab pulsar. Plots show significance, eq.~(\ref{eq:sig}), vs.~injected strength. Color corresponds to the template used for recovery: GR, green; G4v, red; model--independent, blue. This particular search was performed using $10^4$ instantiations, half of which contained injections using the values of $\iota$ and $\psi$ given in table \ref{tab:DetParam}. The model--dependent templates assumed the same same $\iota$ as the injections. Horizontal lines correspond to a detection threshold $\fracdetthrsh=99.9\%$.}
\label{fig:crabS5}
\end{figure*}

We refer to $h_{\rm min}^{\fracdetconf}$ as the \emph{expected sensitivity} or \emph{strain detection threshold at $\fracdetthrsh$}. This value allows not only for the definition of upper limits for the presence of signals, but also the comparison of different model dependent and independent templates. See fig.~\ref{fig:crabS5gr_dep} for a juxtaposition of the results of matching and non--matching model--dependent templates for the case of the Crab pulsar.

The efficiency of a template is also quantified by the slope of the $\hrec$ vs.~$\hinj$ best--fit line, which should be close to 1 for a template that matches the signal. We perform this fit by taking into account only points above the $\fracdetthrsh$ line and forcing the \emph{y}--intersect to be null. The deviations from this fit are used to produce confidence intervals for the recovered strength. This is done by defining a band centered on the best--fit line and enclosing an arbitrary fraction $\fracbandconf$ (e.g.,~$\fracbandconf=95\%$) of the data points, corresponding to the confidence band placed around best--fit line. The intersection between this band and a horizontal line at some value of $\hrec$ yields a confidence interval for the true strength with $\fracbandconf$ confidence. Note that deviations above and below the best--fit line are taken independently to obtain asymmetric confidence intervals. The same analysis can be done on the  $s$ vs.~$\hinj$ plots, taking into account proper scaling of the best--fit slope.

In general, when performing injections we pick parameters with a uniform distribution over the uncertainty ranges of location and orientation values obtained from the ATNF Pulsar Catalog \cite{ATNF}. When there is no orientation information, we must draw $\psi$ and $\iota$ from the ranges $\left[-\pi/2, \pi/2\right]$ and $\left[0,2\pi\right]$ respectively. Note that standard searches consider tensor signals ($2\psi$--dependent) only and therefore assume $\psi \in \left[-\pi/4, \pi/4\right]$; however, a bigger range must be used when taking into account vector signals ($\psi$--dependent). The reason these ranges need not cover the full $\left[0, \pi\right]$ range is that a change in $\psi$ of $\pi/2$ for tensor and $\pi$ for vector signals is equivalent to a change of signal sign. Therefore, this is taken care of by varying the overall phase $\phi_0 \in \left[0, \pi\right]$.

% CRAB PLOTS

%\subsection{Data} \label{sec:data}
We tested the aforementioned methods on LIGO data taken by the Hanford and Livingston detectors over LIGO Science Run 5 (S5). During this run, which took place from November 2005 through September 2007 (GPS times 815155213 - 875232014), the three LIGO detectors operated in data--taking mode at design sensitivity, collecting a year of coincident detector data. The root--mean--square strain noise of the instrument reached values as low as $3\times10^{-22}$ for bands of 100 Hz over the most sensitive frequencies \cite{LSC2009}.
LIGO S5 data has been recently released to the public and is accessible online through the LIGO Open Science Center \cite{losc2014}.

In particular, we looked at data for 115 pulsars, obtained by reducing S5 H1, H2 and L1 strain data as outlined in section \ref{sec:search}. But for the inclusion of PSR J0024-72040 and the exclusion of PSR J2033+17 and Vela, these are the same heterodyned time series analyzed in reference \cite{LSC2010}. However, that study presented Bayesian upper limits to the presence of GR signals and did not consider alternative polarizations.

% MP PLOTS:
\begin{figure*}[!hbtp]
        \centering
        \begin{subfigure}[c]{0.33\textwidth}
                \centering
                \includegraphics[width=\textwidth]{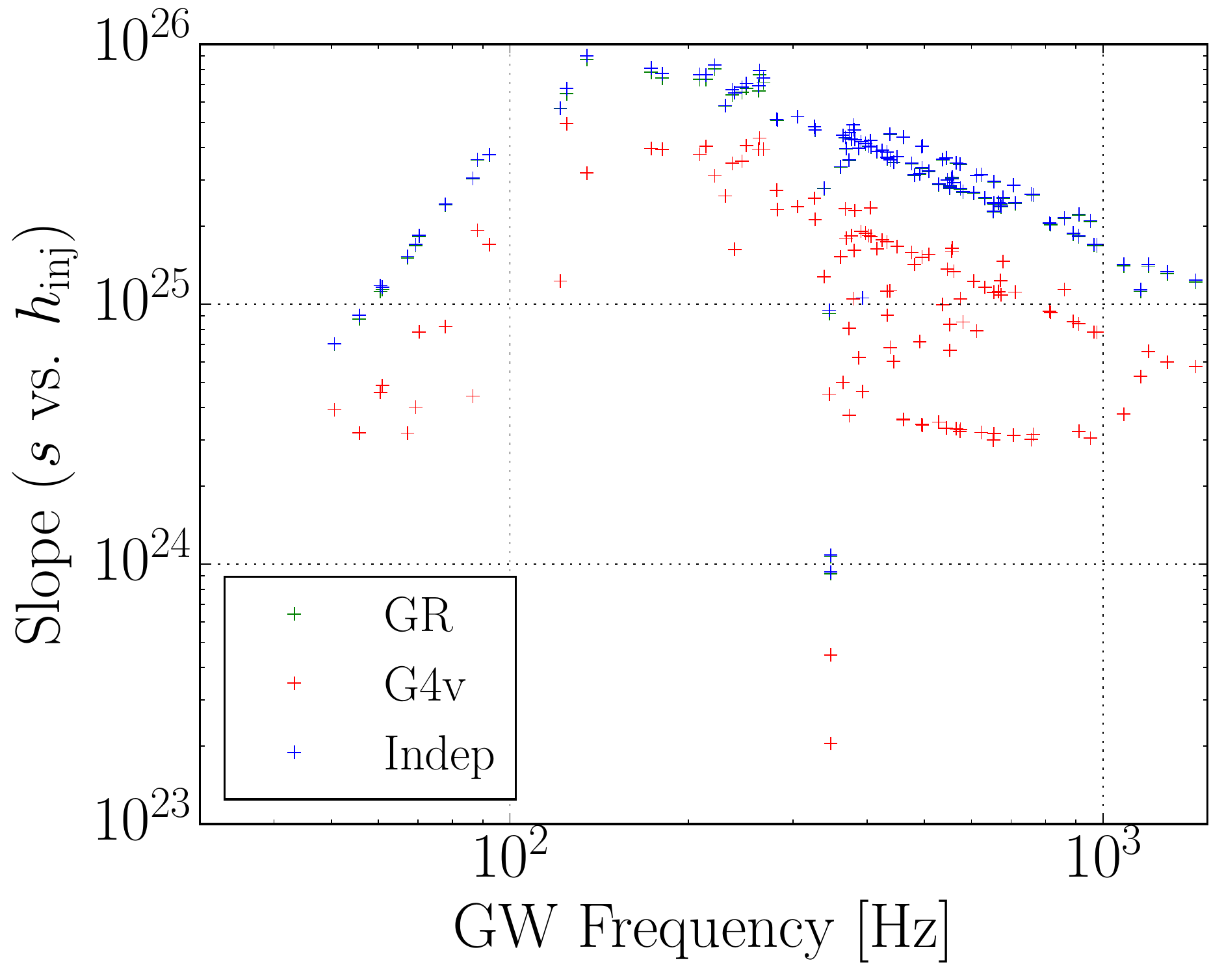}
                \caption{GR slope}
		\label{fig:mpGRslope}
        \end{subfigure}%
        \hfill %
        \begin{subfigure}[c]{0.33\textwidth}
                \centering
                \includegraphics[width=\textwidth]{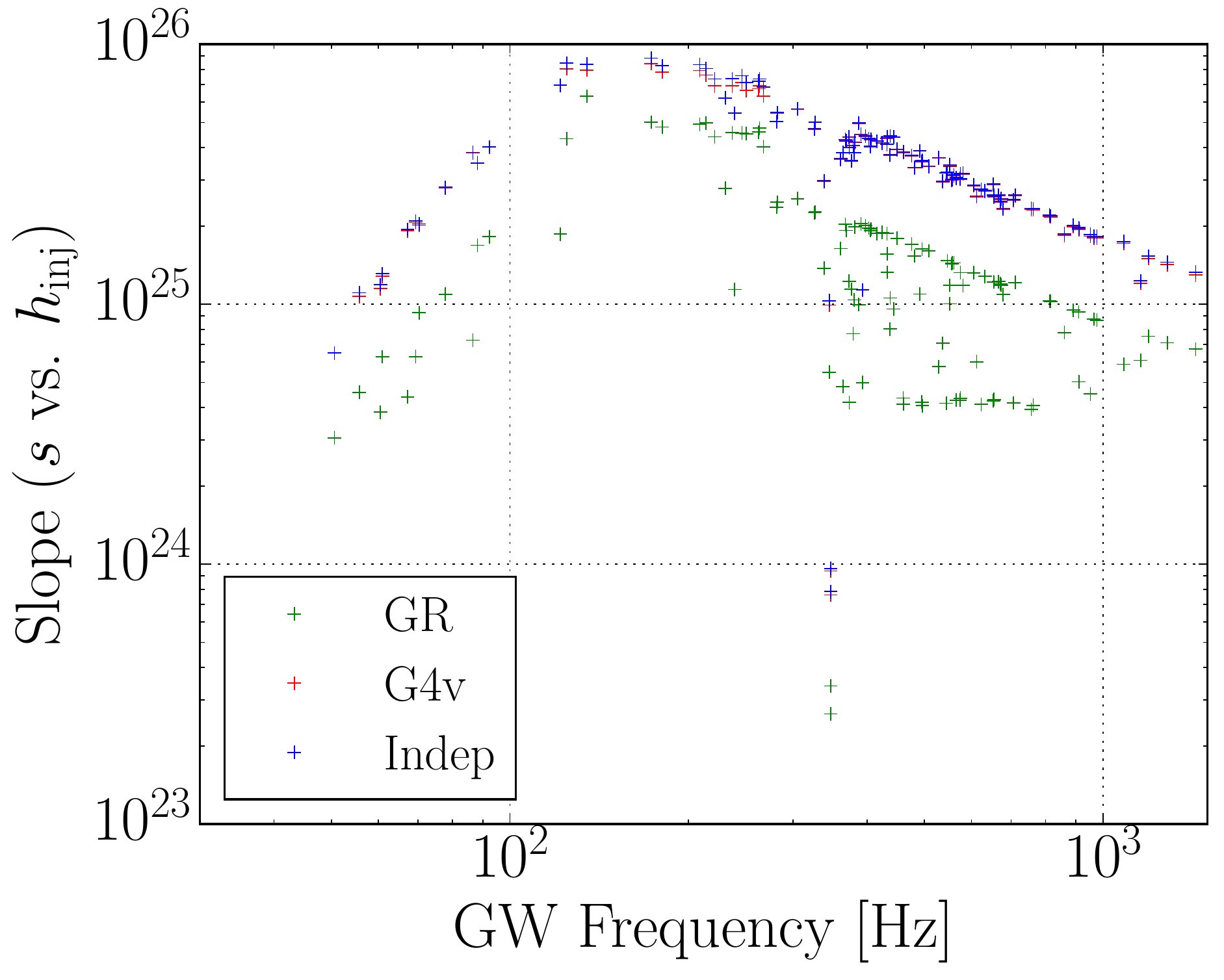}
                \caption{G4v slope}
		\label{fig:mpG4vslope}
        \end{subfigure}%
\hfill
\begin{subfigure}[c]{0.32\textwidth}
                \centering
                \includegraphics[width=\textwidth]{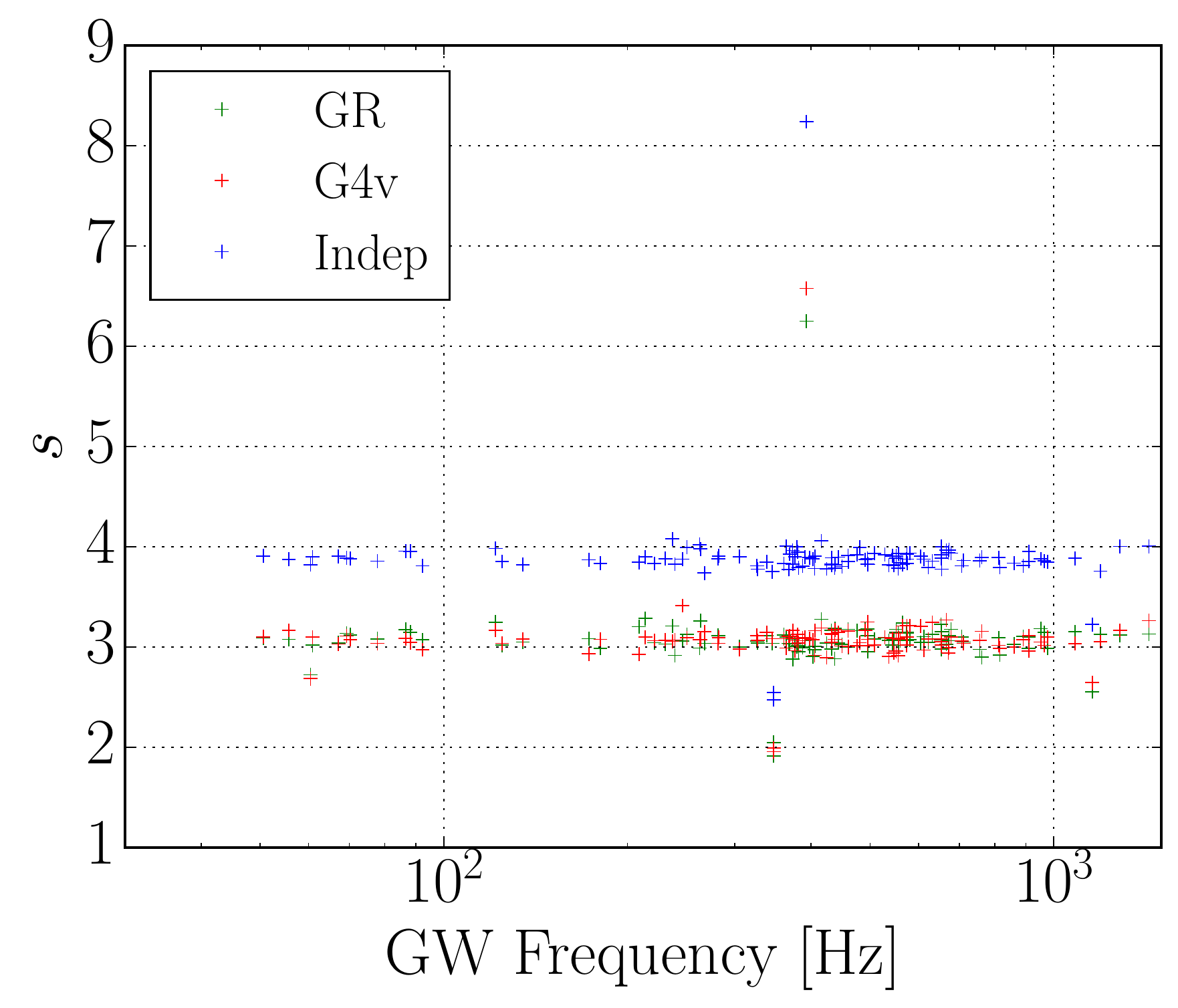}
                \caption{Detection threshold}
		\label{fig:mp_noise}
        \end{subfigure}%
                \caption{Slope of the $s$ vs.~$\hinj$ best--fit--line (left and center) and significance detection threshold at $\fracdetthrsh=99.9\%$ (right) vs.~GW frequency and for GR and G4v injections on S5 H1 data for 115 pulsars. Color corresponds to search template: GR, green; G4v, red; and model--independent, blue. Note that for both kinds of injections, the model--independent points overlap the matching template.}
                \label{fig:mp_slope_noise}
\end{figure*}

\section{Results} \label{sec:results}

Here we present the results of a study of the signal sensitivity of the analysis procedure described in section \ref{sec:analysis}, using the data described at the end of section \label{sec:analysis}. We perform a ``closed box'' analysis, using only re--heterodyned data, which are insensitive to the presence of actual signals, and simulated signal injections. A full ``open box'' analysis, using Bayesian methods to produce model--dependent and model--independent signal detection confidence bands or upper limits, is in preparation.

In particular, we produced $10^4$ re--heterodyned instantiations of data for each pulsar by picking linearly spaced frequencies in the $10^{-7}-10^{-3}$ Hz range (cf.~ sec.~\ref{sec:analysis}). Half of those were injected with simulated signals of increasing strength. The data were then analyzed with each template (GR, G4v and model--independent), producing plots like those in fig.~\ref{fig:injsrch_ex}. For the Crab pulsar, since the source orientation information is known, the full model--dependent templates, eqs.~(\ref{eq:tempGRlocked}, \ref{eq:tempG4vlocked}), were used; otherwise, the semi--model--dependent templates, eqs.~(\ref{eq:tempGR}, \ref{eq:tempG4v}), were used. The whole process was carried out for both GR and G4v injections. In all cases, we set $\fracdetthrsh=99.9\%$ and $\fracdetconf=95.0\%$.

\subsection{Crab pulsar}

Results for searches over H1 S5 data prepared for the Crab pulsar ($\nu=30.22$ Hz, $\nu_{\rm GW}=60.44$ Hz) are presented in fig.~\ref{fig:crabS5}. The results using templates matched to the injections are compared to those of the model--independent (left) and non--matching templates (right). The expected sensitivities, as defined in section \ref{sec:analysis}, for each injection template and search model are provided in table \ref{tab:crabresults}. Recall that the Crab is a special case, since its orientation in the sky is well--known, which enables us to use full model--dependent templates, eqs.~(\ref{eq:tempGRlocked}, \ref{eq:tempG4vlocked}). However, searches for actual signals would still have to make use to two templates for each theoretical model because of the ambiguity in $\iota$ described in section \ref{sec:signal}. In order to avoid doing this, a semi--model--dependent or model--independent search could be carried out instead.

A number of interesting observations can be drawn from fig.~\ref{fig:crabS5} and table \ref{tab:crabresults}. As inferred from the values of $\hmin$, the model--independent template is roughly 25\% less sensitive than the matching one, regardless of the theory assumed when making injections. This is understood by the presence of four extra degrees of freedom in the model--independent template, compared to the single tunable coefficient in the full model--dependent one. If instead the semi--model--dependent template with two degrees of freedom is used, the improvement with respect to the model--independent search goes down to 15\%. In any case, the accuracy of matching and model--independent searches, given by the width of the confidence bands an, are almost identical.

Model dependent templates are significantly less sensitive to non--matching signals. Table \ref{tab:crabresults} indicates that model--dependent templates are 120-170\% less sensitive to non--matching signals than their matching counterpart. A consequence of this is the existence of a range of signals which would be detected by templates of one theory, but not the other (see figs. \ref{fig:crabS5gr_dep} \& \ref{fig:crabS5g4v_dep}). This is particularly interesting, given that previous LIGO searches assume GR to be valid and use a template equivalent to eq.~(\ref{eq:tempGRlocked}). Therefore, our results suggest it is possible that those searches might have missed fully--non--GR signals buried in the data (see section \ref{sec:conclusions} for further discussion).

%% Crab sensitivity [python tab id]
\begin{table}
\centering
      \caption{Summary of expected sensitivity for the Crab pulsar S5 H1 searches ($\fracdetthrsh=99.9\%$, $\fracdetconf=95.0\%$). Rows correspond to injection type and columns to search template. The rotational frequency of the Crab is $\nu=30.22$ Hz and, therefore, $\nu_{\rm GW}=60.44$ Hz.}
\begin{ruledtabular}
\begin{tabular}{lccc}
     & GR           & G4v          & Independent          \\
\cline{2-4}\\[-10pt]
 GR  & \num{3.41e-25} & \num{7.49e-25} & \num{4.20e-25}  \\
 G4v & \num{8.90e-25}  & \num{3.30e-25}  & \num{4.15e-25} \\
\end{tabular}
\end{ruledtabular}
  \label{tab:crabresults}%
\end{table}

% MP results
\begin{figure*}[p]
\begin{subfigure}[b]{\textwidth}
                \includegraphics[width=.7\textwidth]{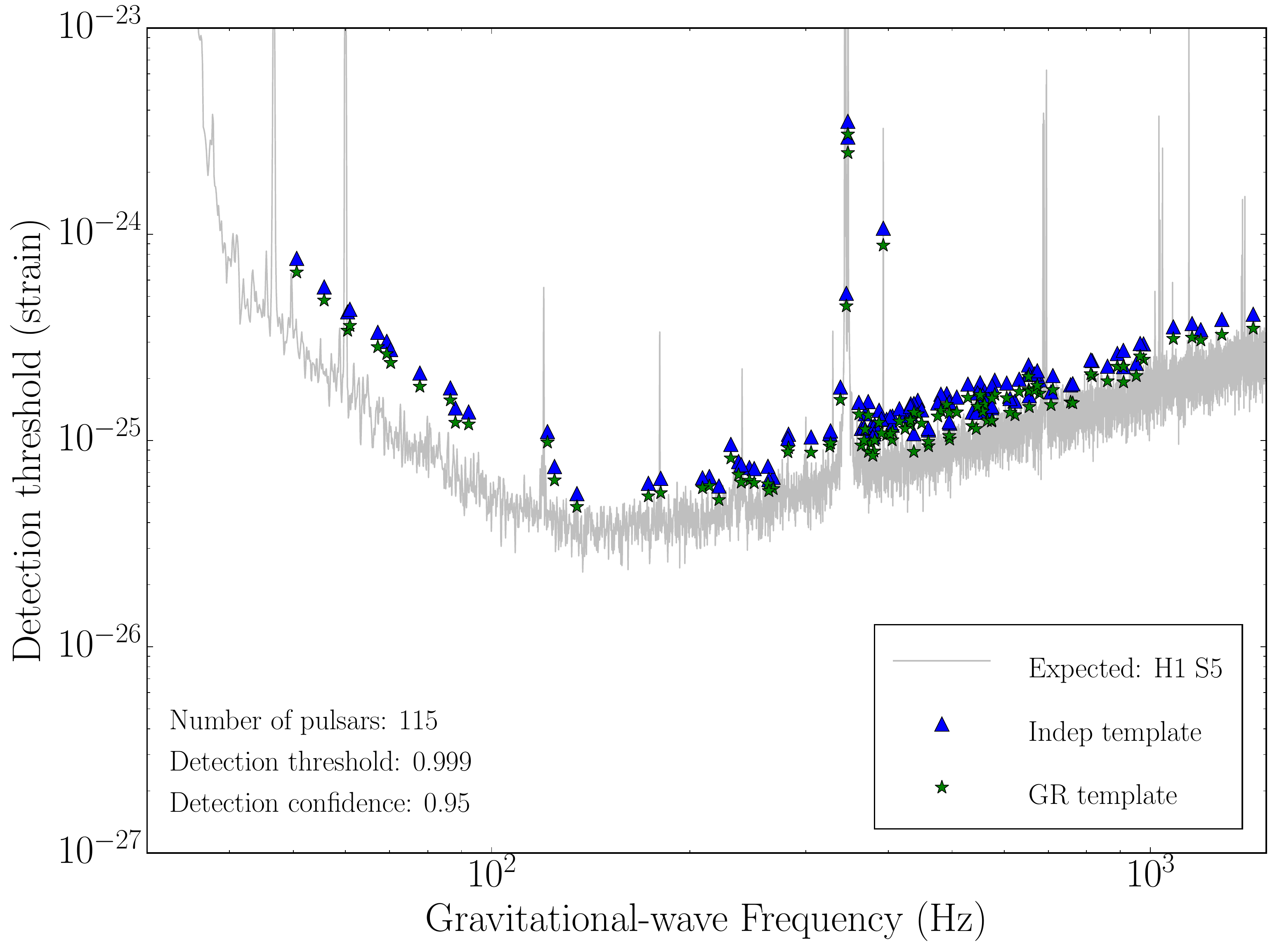}
        \caption{GR injections}
\label{fig:mpGR}
\end{subfigure}\\
\begin{subfigure}[b]{\textwidth}
                \includegraphics[width=.7\textwidth]{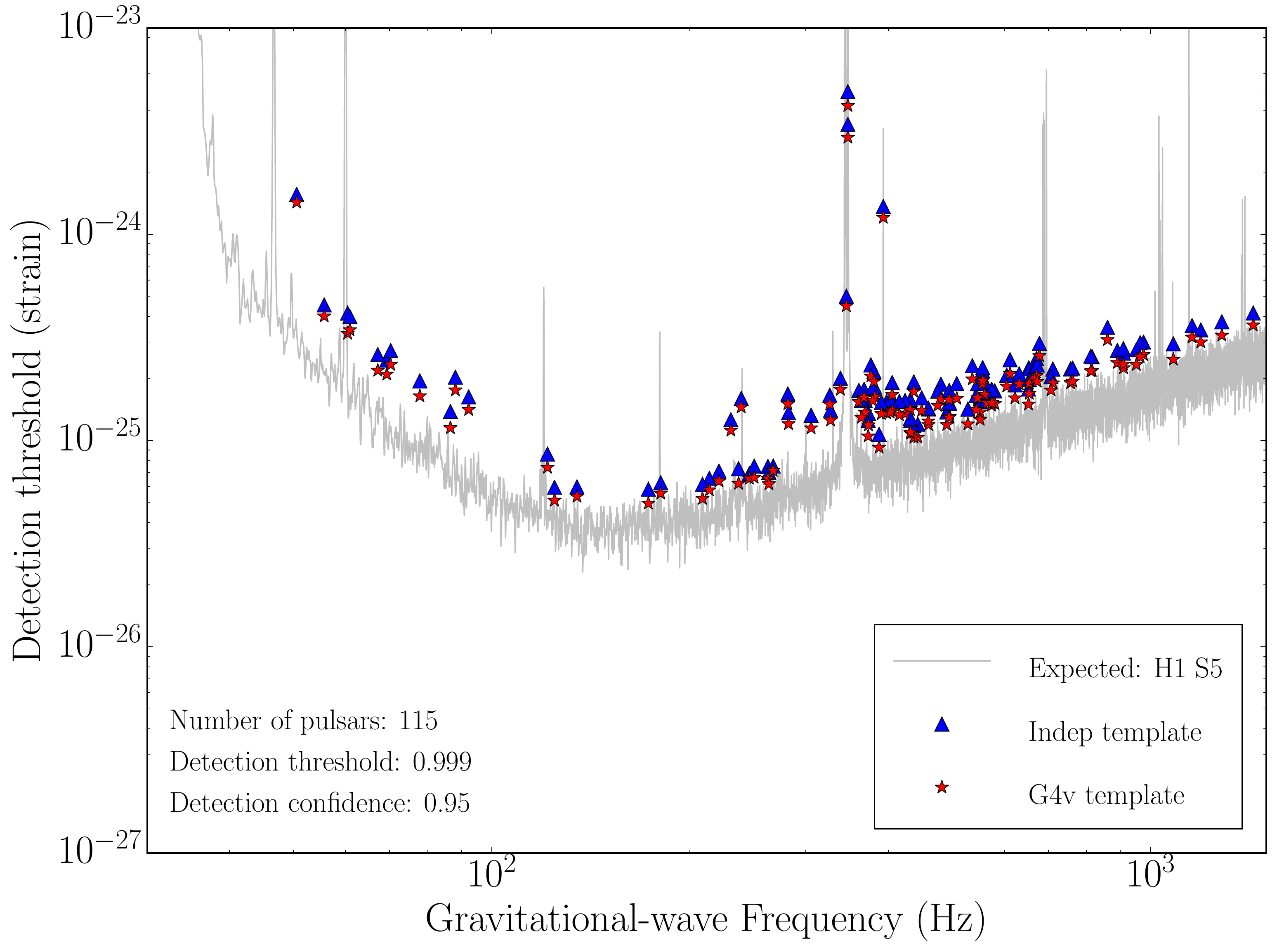}
        \caption{G4v injections}
\label{fig:mpG4v}
\end{subfigure}%
\caption{S5 H1 expected sensitivity (strain detection threshold at $\fracdetthrsh=99.9\%$ with $\fracdetconf=95.0\%$ confidence) vs.~GW frequency for 115 pulsars. Color corresponds to search template: GR, green; G4v, red; and model--independent, blue. The gray line is the anticipated sensitivity of a standard Bayesian search, eq.~(\ref{eq:expsens}).}
\label{fig:mpH1}
%\vspace{-15pt}
\end{figure*}

\begin{figure*}[p]
\begin{subfigure}{\textwidth}
                \centering
                \includegraphics[width=.7\textwidth]{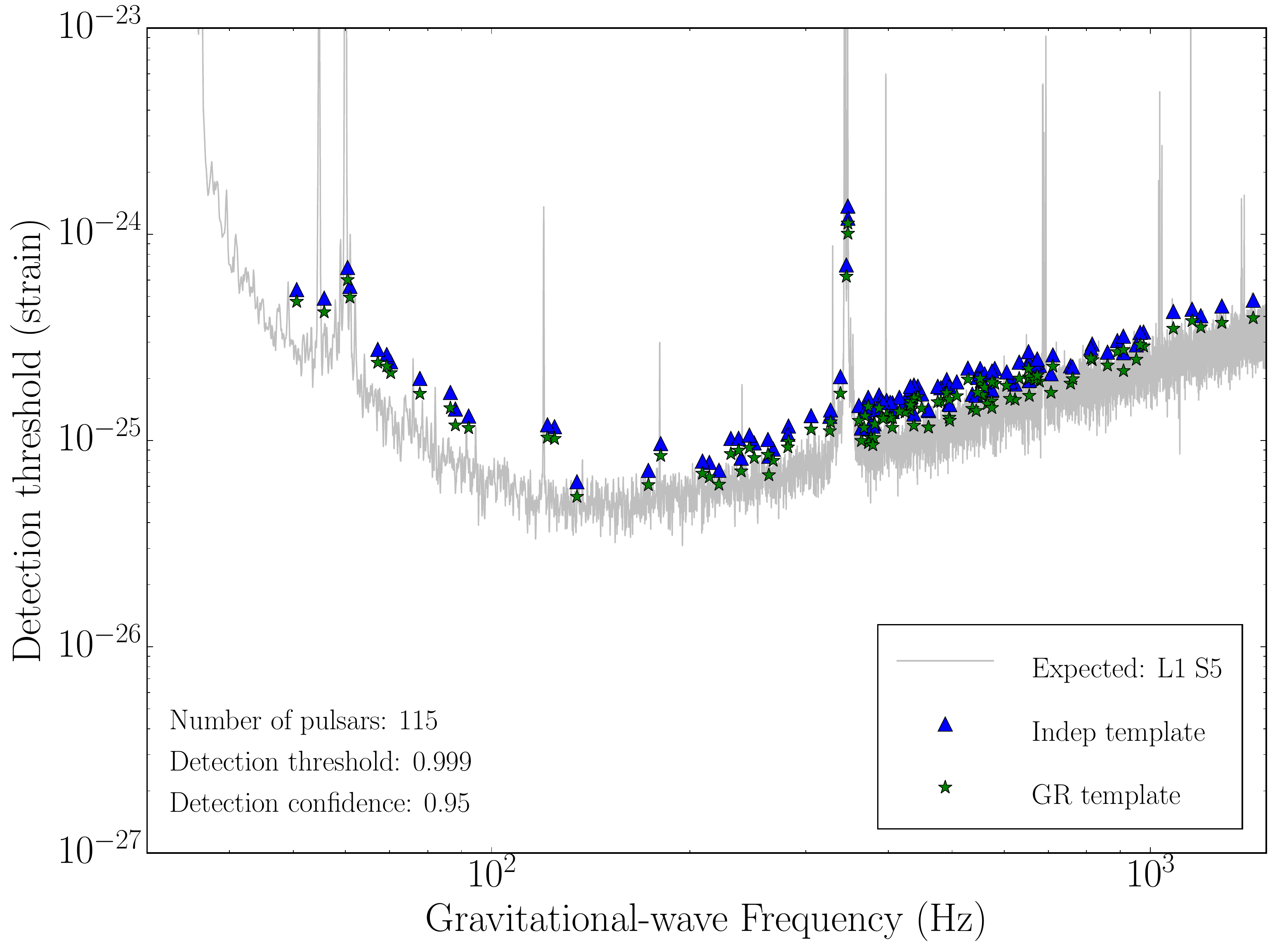}
        \caption{GR injections}
\label{fig:mpGR}
\end{subfigure}%
\\
\begin{subfigure}{\textwidth}
\includegraphics[width=.7\textwidth]{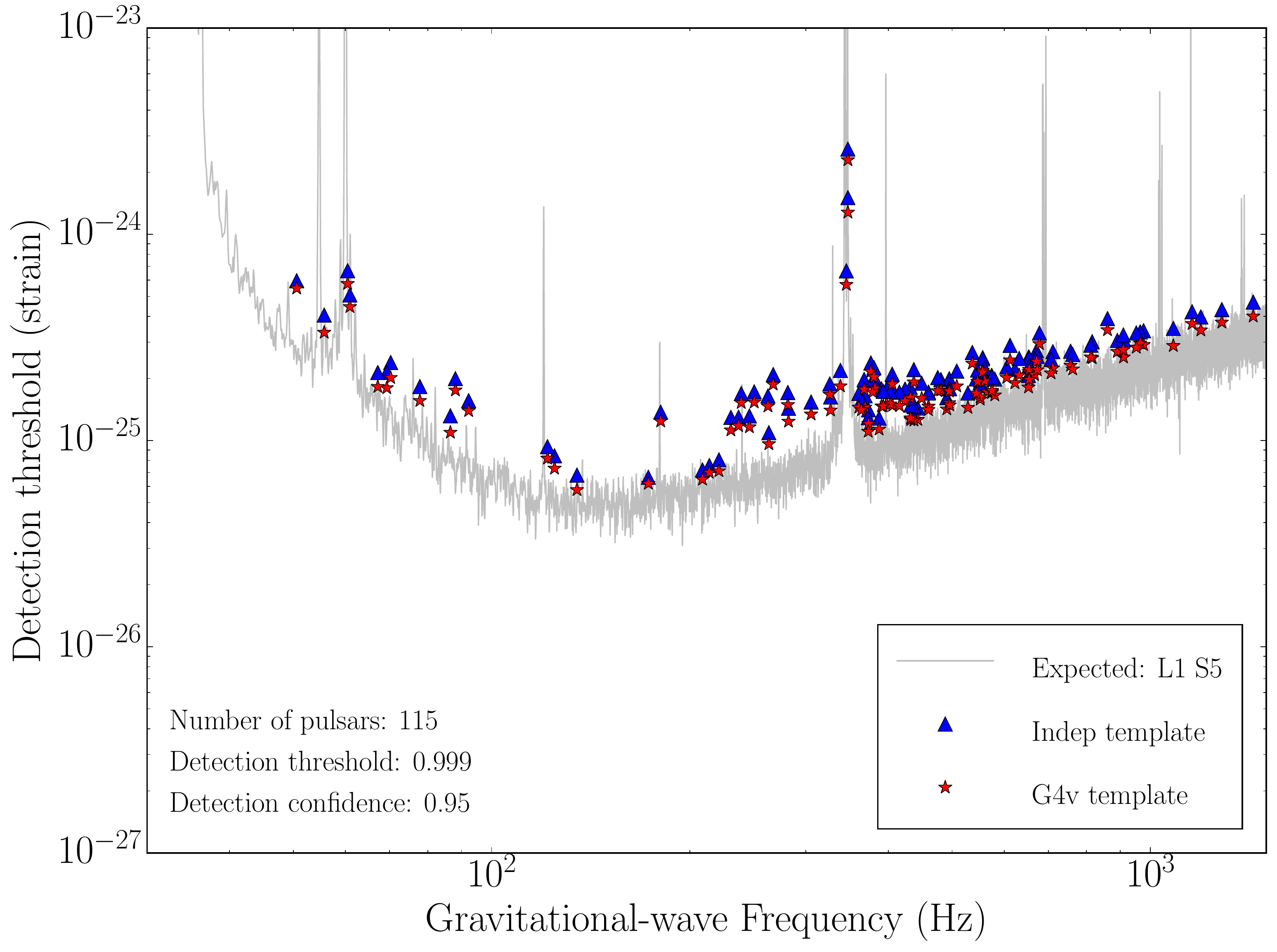}
\caption{G4v injections}
\label{fig:mpG4v}
\end{subfigure}%
\caption{S5 L1 expected sensitivity (strain detection threshold at $\fracdetthrsh=99.9\%$ with $\fracdetconf=95.0\%$ confidence) vs.~GW frequency for 115 pulsars. Color corresponds to search template: GR, green; G4v, red; and model--independent, blue. The gray line is the anticipated sensitivity of a standard Bayesian search, eq.~(\ref{eq:expsens}).}
\label{fig:mpL1}
%\vspace{-15pt}
\end{figure*}

\subsection{All pulsars}

The Crab pulsar is only one of the 115 sources we analyzed. The results, presented in figs.~\ref{fig:mpGRslope} \& \ref{fig:mpG4vslope} generally confirm the observations anticipated from the Crab. While model--independent searches are of the same accuracy as matching semi--model--dependent ones, their strain detection threshold is louder due to the extra degrees of freedom (fig.~\ref{fig:mp_noise}). Consequently, model--independent templates demand a higher significance to be able to distinguish a signal from noise. The detection thresholds for GR and G4v templates are of the same magnitude, since both have the same number of degrees of freedom. Among all the 115 pulsars, the sources with best expected sensitivities to GR and G4v signals were PSR J1603-7202 and PSR J1748-2446A respectively (see table \ref{tab:bestresults}).

%%H1S5 sensitivity [python tab id]
\begin{table}
\centering
      \caption{Best expected sensitivities for S5 H1 searches ($\fracdetthrsh=99.9\%$, $\fracdetconf=95.0\%$). Rows correspond to injection type and columns to pulsar name (PSR), rotation frequency ($\nu$) and strain detection threshold for matching dependent ($h_{\rm dep}$) and independent ($h_{\rm indep}$) templates.}
\begin{ruledtabular}
\begin{tabular}{ccccc}
  & PSR        & $\nu$ (Hz)       & $h_{\rm dep}$   & $h_{\rm indep}$   \\
\cline{2-5}\\[-10pt]
 GR  & J1603-7202  & \num{67.38} & \num{4.77e-26}  & \num{5.53e-26}    \\
 G4v & J1748-2446A & \num{86.48} & \num{4.96e-26}  & \num{5.81e-26}    \\
\end{tabular}
\end{ruledtabular}
\label{tab:bestresults}
\end{table}

The key results of our study are summarized in fig.~\ref{fig:mpH1} for H1 and fig.~\ref{fig:mpL1} for L1. These plots present the expected sensitivity (strain detection threshold at $\fracdetthrsh=99.9\%$ with $\fracdetconf=95.0\%$ confidence) vs.~GW frequency ($\nu_{\rm GW}=2\nu$). The outliers seen in figs.~\ref{fig:mp_slope_noise}-\ref{fig:mpL1} correspond to pulsars whose value of $\nu_{\rm GW}$ are very close to instrumental noise spectral lines associated with violin resonances of the detectors’ test mass pendulum suspensions.

For the matching or model--independent templates, the resulting data points trace the noise curve of the instrument; however, due to the long integration time, we are able to detect signals below LIGO's standard strain noise. The gray curve shown in figs.~\ref{fig:mpH1}, \ref{fig:mpL1} represents the expected sensitivity of a regular Bayesian GR search (e.g.,~\cite{LSC2010}). This is proportional to the amplitude spectral density of the detector and inversely proportional to the square--root of the observation time. The particular empirical relationship used to generate the curve in figs.~\ref{fig:mpH1} \& \ref{fig:mpL1} is:
\beq \label{eq:expsens}
\langle \hmin \rangle = 10.8 \sqrt{S_n(f)/T},
\eeq
with $S_n(f)$ the noise power spectral density and $T$ the total observation time (527 days for S5 H1 and 405 days for S5 L1) \cite{Dupuis2005}. This formula enables the comparison of the methods presented here with the expected performance of standard Bayesian searches.

\begin{figure}[!hbtp]
        \centering
        \begin{subfigure}[c]{0.24\textwidth}
                \centering
                \includegraphics[width=\textwidth]{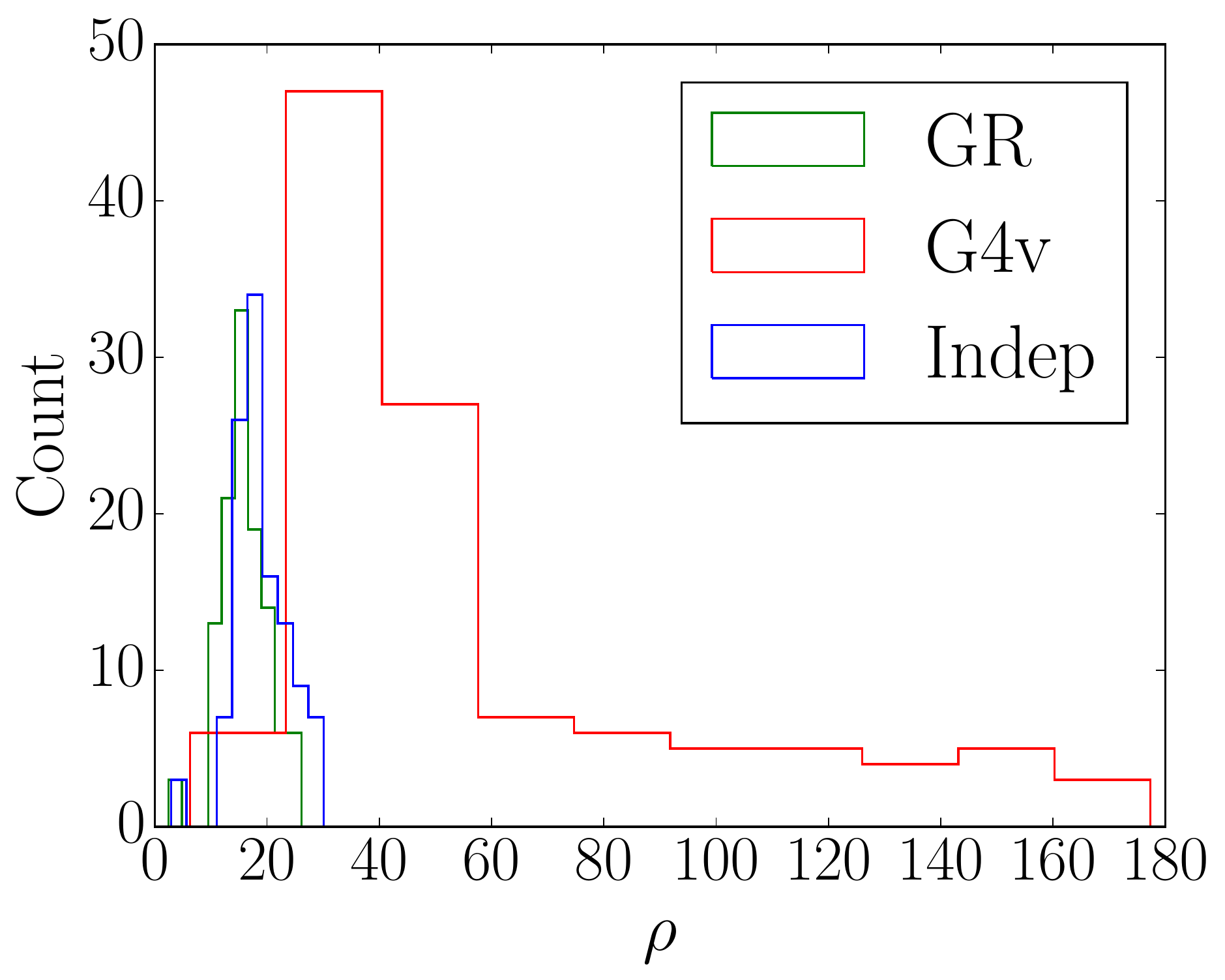}
                \caption{H1 GR}
		\label{fig:mp_rho_H1GR}
        \end{subfigure}%
        \hfill %
        \begin{subfigure}[c]{0.24\textwidth}
                \centering
                \includegraphics[width=\textwidth]{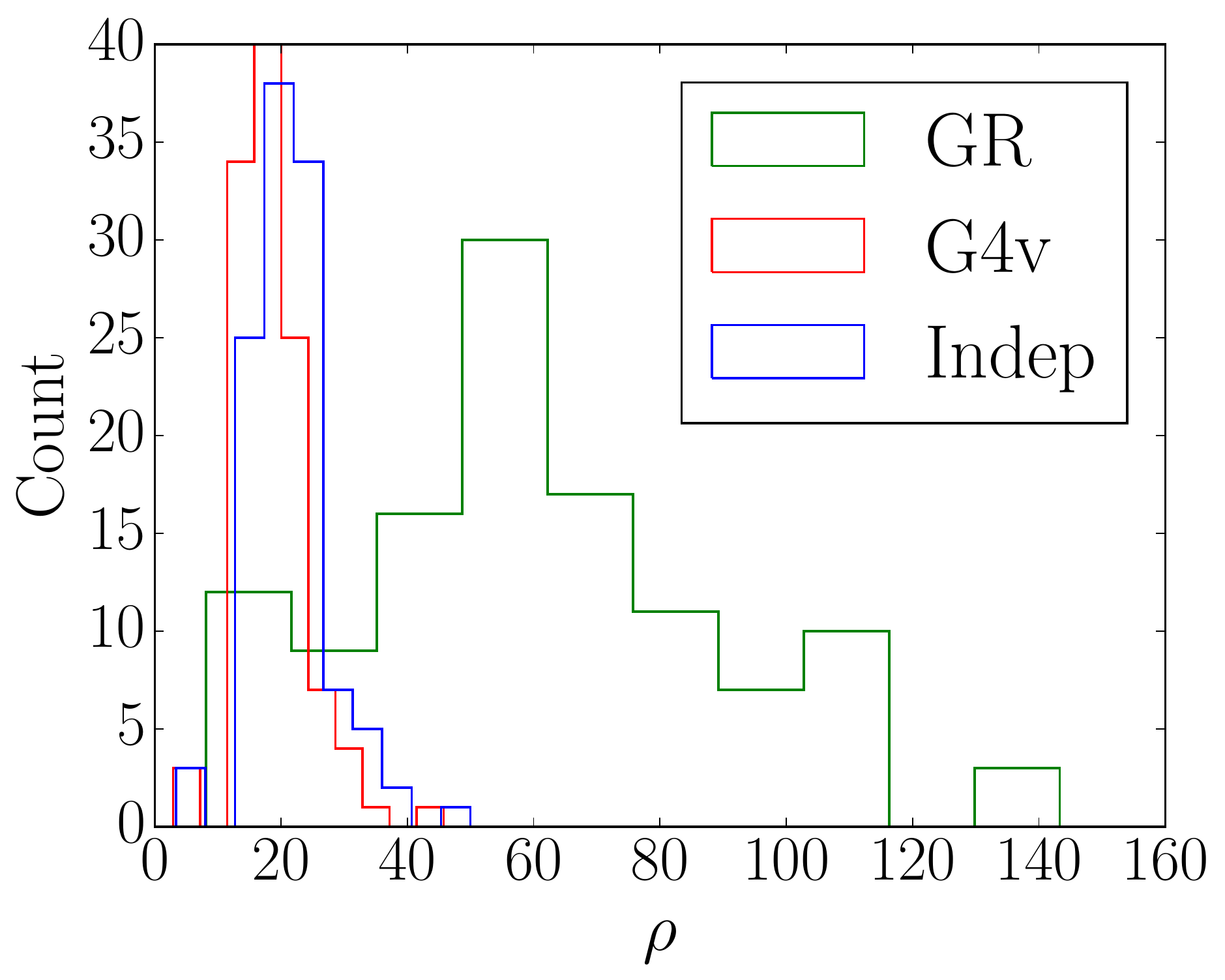}
                \caption{H1 G4v}
		\label{fig:mp_rho_H1G4v}
        \end{subfigure}%
        \hfill %
        \begin{subfigure}[c]{0.24\textwidth}
                \centering
                \includegraphics[width=\textwidth]{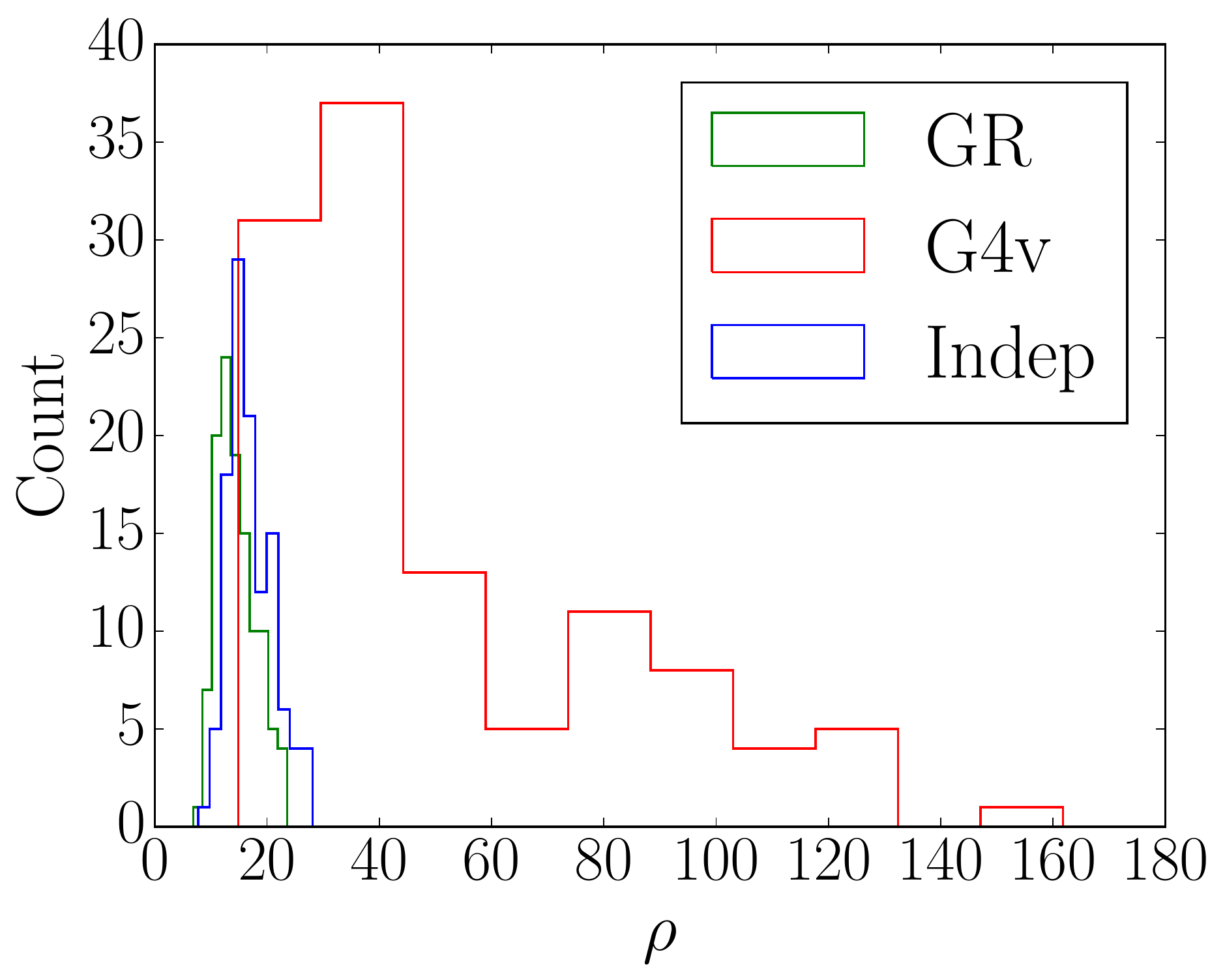}
                \caption{L1 GR}
		\label{fig:mp_rho_L1GR}
        \end{subfigure}%
        \hfill %
        \begin{subfigure}[c]{0.24\textwidth}
                \centering
                \includegraphics[width=\textwidth]{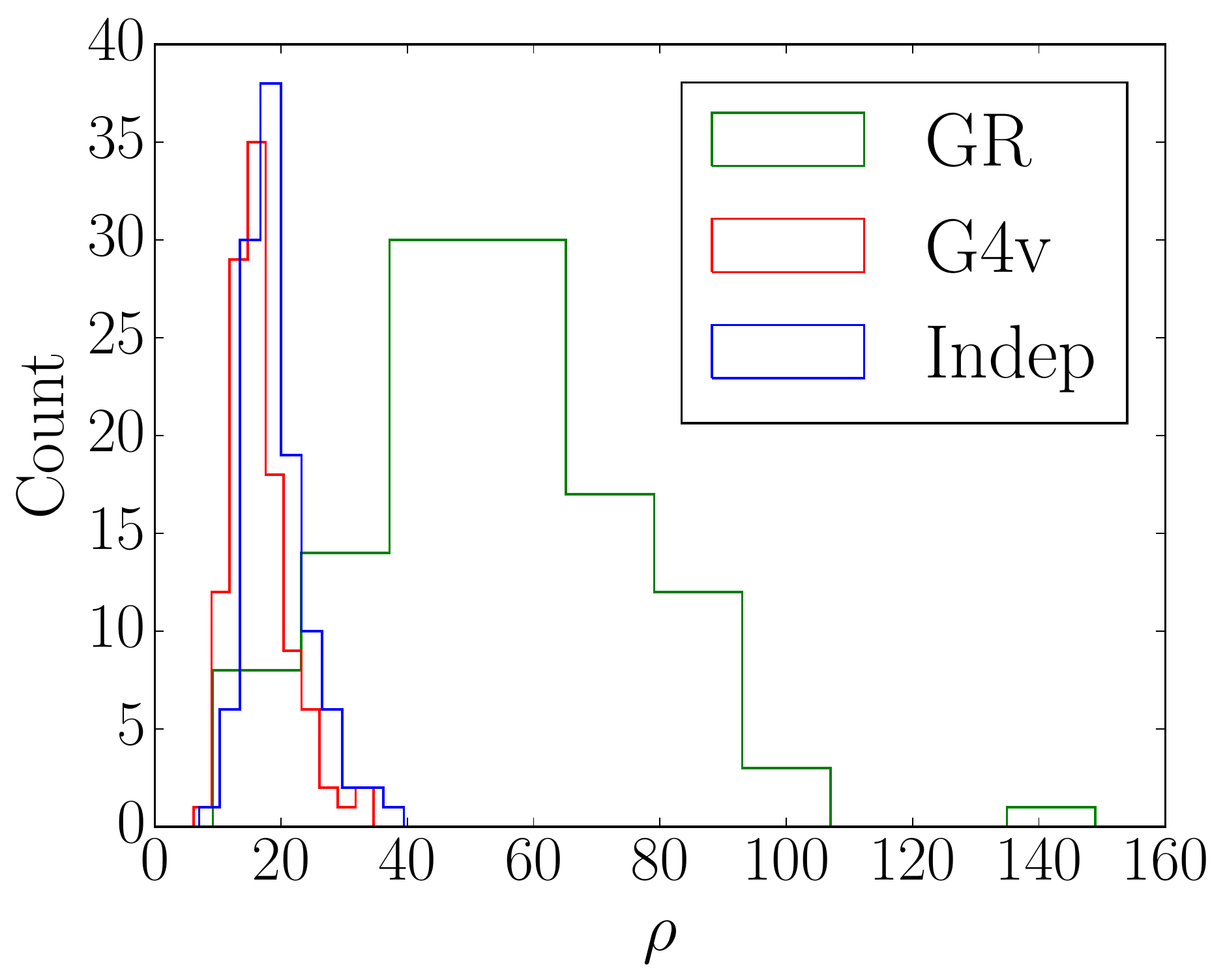}
                \caption{L1 G4v}
		\label{fig:mp_rho_L1G4v}
        \end{subfigure}
                \caption{Histograms of the figure of merit $\rho$, eq.~(\ref{eq:rho}), for our searches over S5 H1 (top) and L1 (bottom) data sets with GR (left) and G4v (right) injections, corresponding to 115 pulsars. Color corresponds to search template: GR, green; G4v, red; and model--independent, blue.}
                \label{fig:mp_rho}
\end{figure}

By the same token, we can define a figure of merit $\rho$ for our searches by the ratio:
\beq \label{eq:rho}
\rho\left(\nu_{\rm GW}\right) = \hmin / \sqrt{S_n(\nu_{\rm GW})/T}.
\eeq
The average of this value over all pulsars, $\langle\rho\rangle$, can be semi--quantitatively compared to the 10.8 prefactor in eq.~(\ref{eq:expsens}). The equivalence is not direct because, besides the intrinsic differences between Bayesian and frequentist approaches, eq.~(\ref{eq:expsens}) was obtained by averaging the results of 4000 simulated searches \cite{Dupuis2005}, while we include just the 115 pulsars at hand. The values of $\langle\rho\rangle$ for our S5 H1 \& L1 analyses are presented in table \ref{tab:sensratios} and fig.~\ref{fig:mp_rho}. The specific values for the Crab pulsar are shown in table \ref{tab:crabsensratios}. A smaller $\rho$ indicates better performance.

%%H1S5 avg rho [python tab id]
\begin{table}
\centering
      \caption{Average sensitivity ratios $\langle\rho\rangle$, eq.~(\ref{eq:rho}), for S5 H1 (first value) and S5 L1 (second value) searches.  Rows correspond to injection type and columns to search template.}
\begin{ruledtabular}
\begin{tabular}{cccc}
  & GR        & G4v       & Independent   \\
\cline{2-4}\\[-10pt]
 GR  & 16.11 ~~~ 14.65 & 58.53 ~~~ 51.89 & 18.83 ~~~ 17.15\\
 G4v & 61.21 ~~~ 55.06 & 18.42 ~~~ 16.76 & 21.24 ~~~ 19.32 \\
\end{tabular}
\end{ruledtabular}
\label{tab:sensratios}
\end{table}

%%H1S5 Crab rho [python tab id]
\begin{table}[t]
      \caption{Crab sensitivity ratio $\rho$, eq.~(\ref{eq:rho}) evaluated at the Crab's GW frequency, for S5 H1 (first value) and S5 L1 (second value) searches.  Rows correspond to injection type and columns to search template.}
\begin{ruledtabular}
\begin{tabular}{cccc}
  & GR        & G4v       & Independent   \\
\cline{2-4}\\[-10pt]
 GR  & 20.75 ~~~ 10.40 & 45.52 ~~~ 27.15& 25.54 ~~~ 11.94 \\
 G4v & 54.06 ~~~ 20.30 & 20.07 ~~~ 9.96 & 25.21 ~~~ 11.52\\
\end{tabular}
\end{ruledtabular}
\label{tab:crabsensratios}
\end{table}

As mentioned above, the remarks made about the Crab pulsar hold for most other sources, except that detectability is slightly lower because orientation parameters are unknown. In all cases, the matching template is the best at recovering signals, followed closely by the model--independent one. Searches that assume the incorrect model are substantially less efficient and their $\hmin$ vs.~$\nu_{\rm GM}$ curves do not follow the instrumental noise line. This is reflected, for instance, by the figures of merit presented in table \ref{tab:sensratios}.

\section{Conclusions}  \label{sec:conclusions}

We have developed novel model--independent methods to search for CGW signals coming from targeted sources in LIGO--like interferometric data. These searches are able to detect signals of \emph{any} polarization content with high significance.

In order to test our methods in the presence of realistic noise conditions, we implemented a procedure to produce thousands of noise--only instantiations from actual data. We then proceeded by injecting and retrieving increasingly loud signals of different polarization content.

We studied 115 pulsars using S5 data from the LIGO Hanford and Livingston detectors. Although the methods are general, we restricted our study to two theories that predict starkly different GW polarization contents (GR and G4v).

Our results indicate that assuming the wrong theoretical model greatly reduces the sensitivity of a search to signals buried in the data. Yet, our model--independent searches are almost as effective as the model--dependent templates that match the kind of signal injected (i.e.~when the models used for injection and search are the same). This means that our model--independent templates can be used to find signals of any polarizations without additional computational requirements.

We are able to reach sensitivities comparable to previous studies, although slightly worse than those presented in \cite{LSC2010}. This is probably due to our making use of a single detector and to differences between frequentist and Bayesian approaches.

We have shown that, for some combinations of detectors, sources, and signal strengths, G4v signals are invisible to GR templates and vice--versa. Therefore, it is possible that, if GWs are composed uniquely of vector modes, previous LIGO searches, which assume GR, may have missed their signals.

It is clear that the next step in this study consists of incorporating our model--independent templates into the Bayeasian machinery used in standard LIGO Scientific Collaboration searches. This will allow us to properly marginalize over all nuisance parameters and to produce multi–-detector model--dependent and model--independent signal detection confidence bands or upper limits. We will also employ methods to constrain other theories (e.g.,~scalar--tensor) in the event of a model-independent detection.

\begin{acknowledgments}
The authors would like to thank Holger Pletsch for helpful discussions. M.~Pitkin is funded by the STFC through grant number ST/L000946/1. LIGO was constructed by the California Institute of Technology and Massachusetts Institute of Technology with funding from the National Science Foundation and operates under cooperative agreement PHY-0757058. This paper carries LIGO Document Number LIGO-P1400169.
\end{acknowledgments}

\appendix
\section{Statistical properties of LIGO data} \label{sec:stats}

The $\chi^2$ minimization is equivalent to a maximum likelihood procedure only in the presence of Gaussian noise. When this requirement is not satisfied, the regression is still valid, but the $\chi^2$ values resulting from the fit will be distributed in a non--trivial way, rather than the $\chi^2$ distribution expected in the case of Gaussian noise. Furthermore, the relationship between the covariance matrix of the system and the standard uncertainties of the recovered coefficients becomes unclear. Therefore, it is important to statistically characterize the data and understand the limitations of our assumption of Gaussianity.

\begin{figure*} [!hbt]
\begin{subfigure}[b]{\columnwidth}
\includegraphics[width=\textwidth]{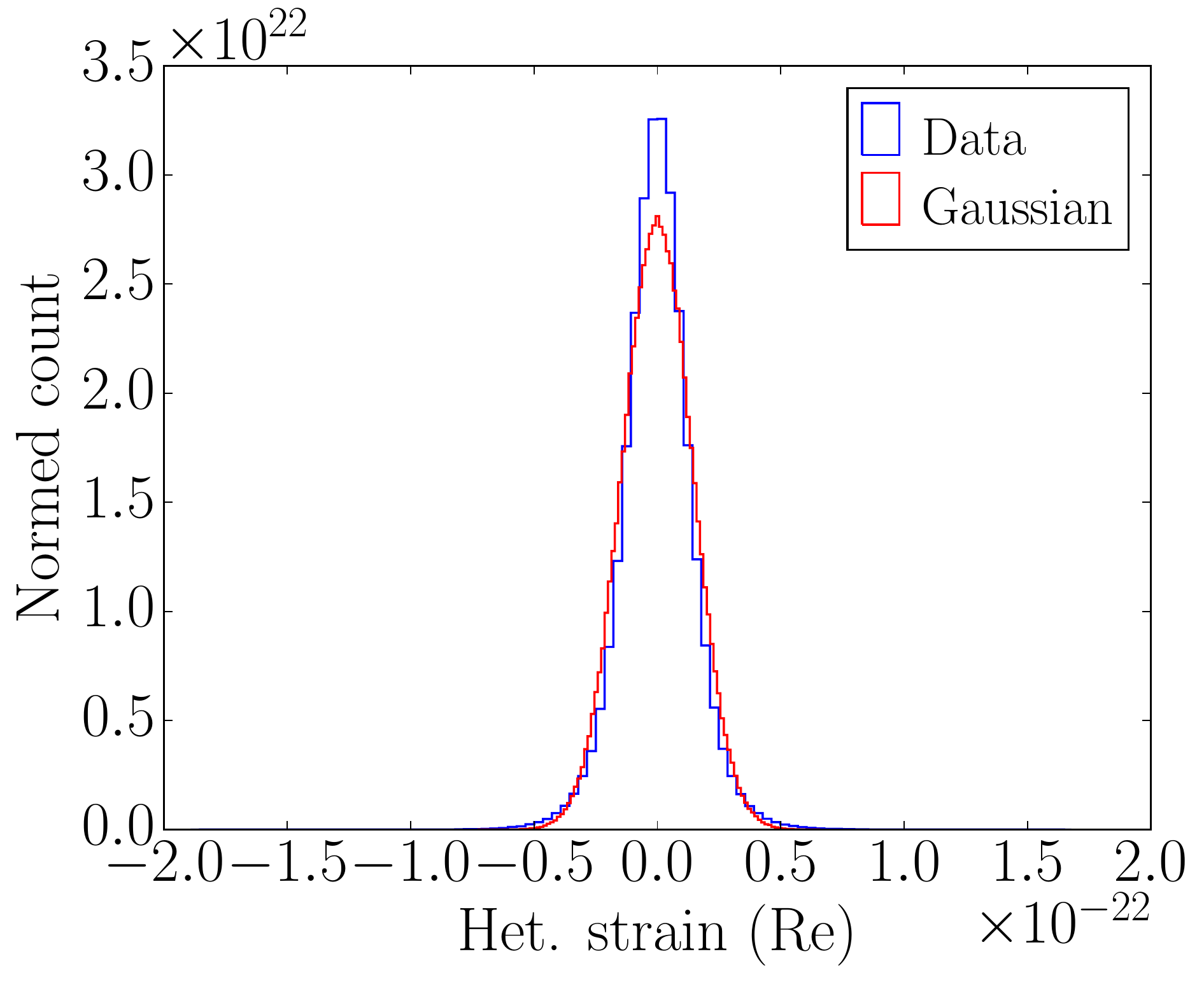}
%\caption{H1 S5 Crab histogram}
\end{subfigure}\hfill
\begin{subfigure}[b]{\columnwidth}
\includegraphics[width=\textwidth]{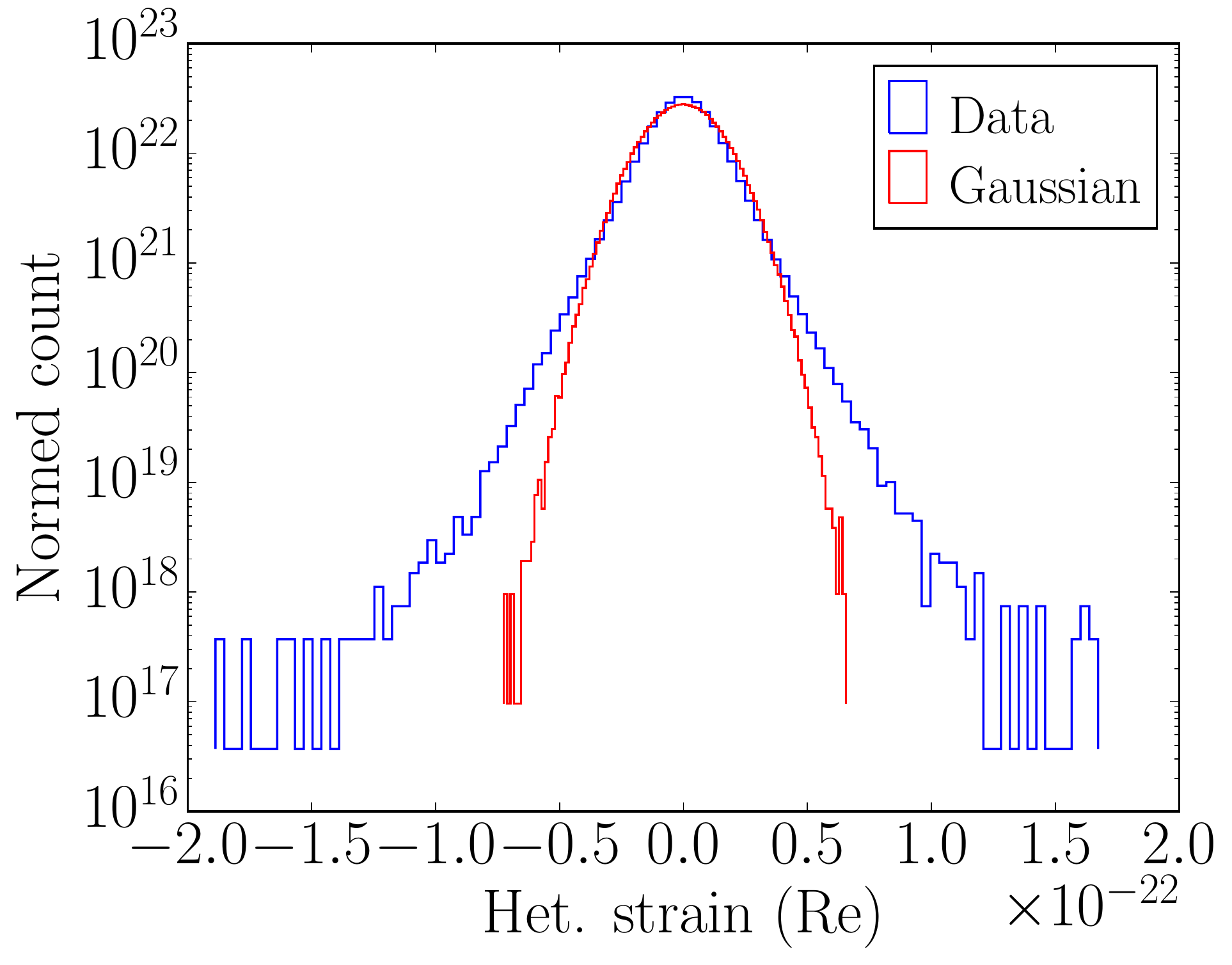}
%\caption{H1 S5 Crab histogram logscale}
\end{subfigure}%
\caption{Normalized histogram of the real part of S5 H1 data heterodyned for the Crab in linear (left) and logarithmic $y$ scales. A Gaussian curve with the same standard deviation is plotted in red for comparison.}
\label{fig:crabhist}
\end{figure*}

When taken as a whole, LIGO detector noise does not conform to a \emph{stationary} Gaussian distribution. This can be visually confirmed by means of a histogram, as shown in fig.~\ref{fig:crabhist} for the case of S5 H1 data prepared for the Crab. The divergence from Gaussianity is evident from the long tails, seen most clearly in the log--$y$ version of the plot. As expected, the data fail more rigorous standard Gaussianity tests, such as the Kolmogorov--Smirnov (KS) or the Anderson--Darling (AD) tests.

\begin{figure*} [!hbt]
\begin{subfigure}[t]{\columnwidth}
\includegraphics[width=\textwidth]{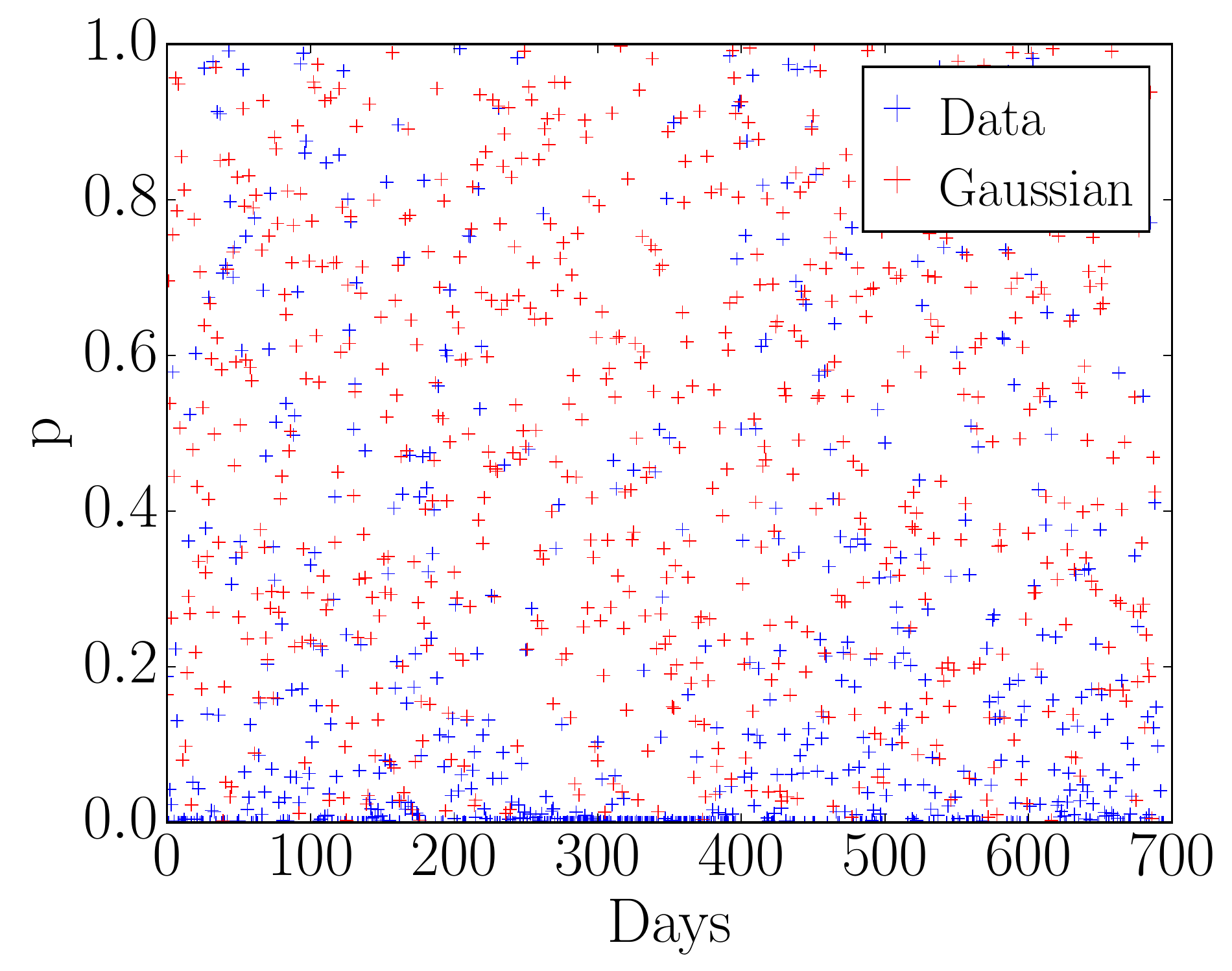}
\caption{Kolmogorov--Smirnov}
\label{fig:kstest}
\end{subfigure}\hfill
\begin{subfigure}[t]{\columnwidth}
\includegraphics[width=\textwidth]{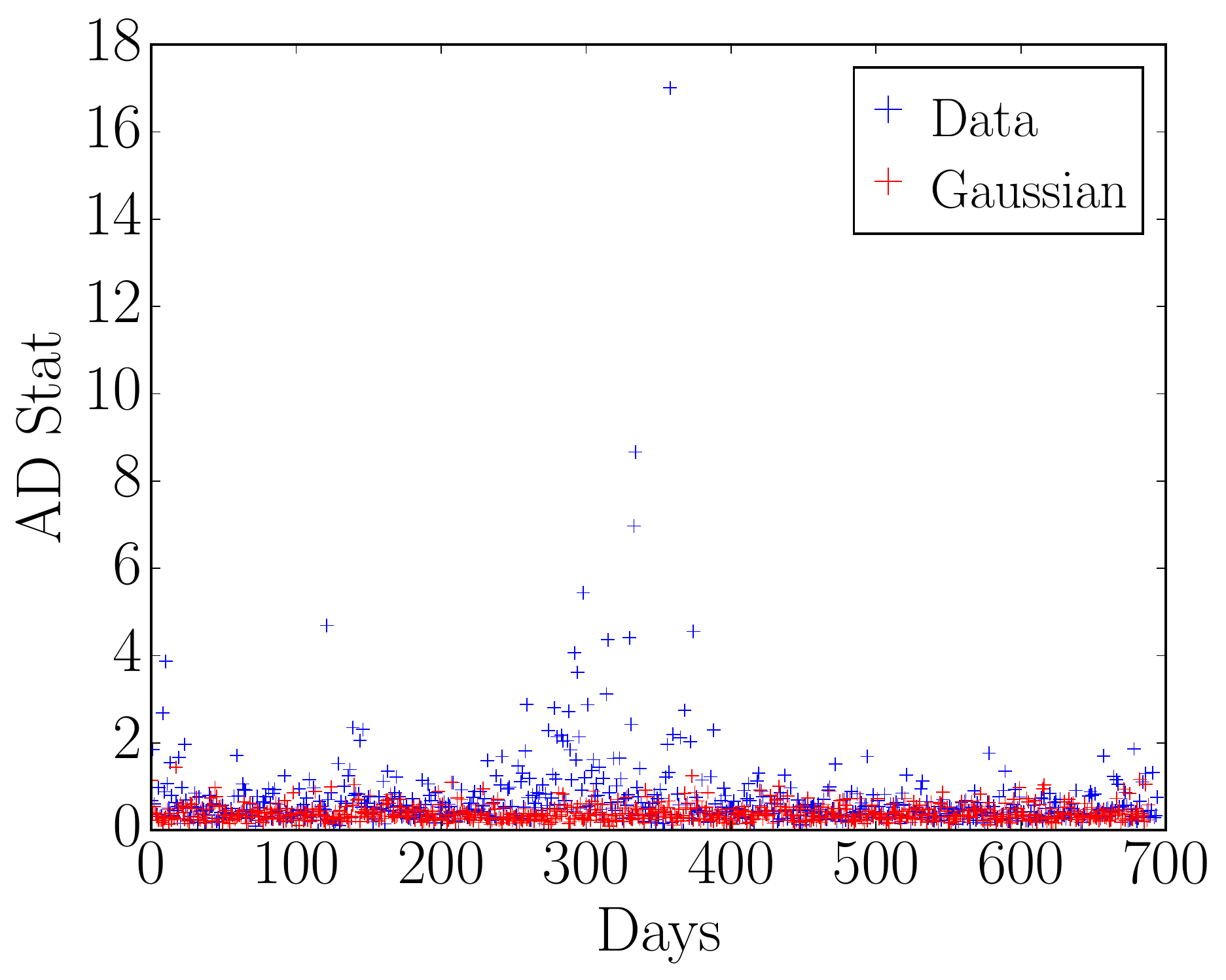}
\caption{Anderson--Darling}
\label{fig:adtest}
\end{subfigure}%
\caption{Results of Gaussianity tests for daily segments of S5 H1 data prepared for the Crab (blue). The results for corresponding sets of Gaussian noise are presented for comparison (red).}
\label{fig:gausstests}
\end{figure*}

However, it is possible to split up the data into day--long (or shorter) segments, as was described in section \ref{sec:search}, so as to study the Gaussianity of the data on a day--to--day basis. The results of the KS and AD tests for each day--segment, together with those for reference Gaussian noise series, are presented in figs.~\ref{fig:kstest} and \ref{fig:adtest} respectively. The KS test returns the $p$--value for a null hypothesis that assumes the data is normally distributed; therefore, a lower $p$--value implies a higher probability that the data are not Gaussian \cite{chakravarty1967}. The AS test returns a figure of merit which is indirectly proportional to the significance with which the hypothesis of Gaussianity can be rejected; therefore a higher AS statistic implies a higher probability that the data are not Gaussian \cite{Anderson1952}.


It can be seen from the results of these tests that the statistical properties of the segments vary considerably from day to day. This could have been guessed from the non-stationarity of the data in fig.~\ref{fig:finehetS5reH1}, the daily variation of the standard deviation (fig.~\ref{fig:std}) and other irregularities of the data. Nonetheless, most of the segments seem to pass the Gaussianity tests, with some remarkable exceptions around the days 250--400 of the run. This corresponds to the spiking observed in the heterodyned data (GPS times $8.4\times10^8-8.5\times10^8$ in fig.~\ref{fig:finehetS5reH1}).

\begin{figure*}[p]
\begin{subfigure}{\textwidth}
                \includegraphics[width=.7\textwidth]{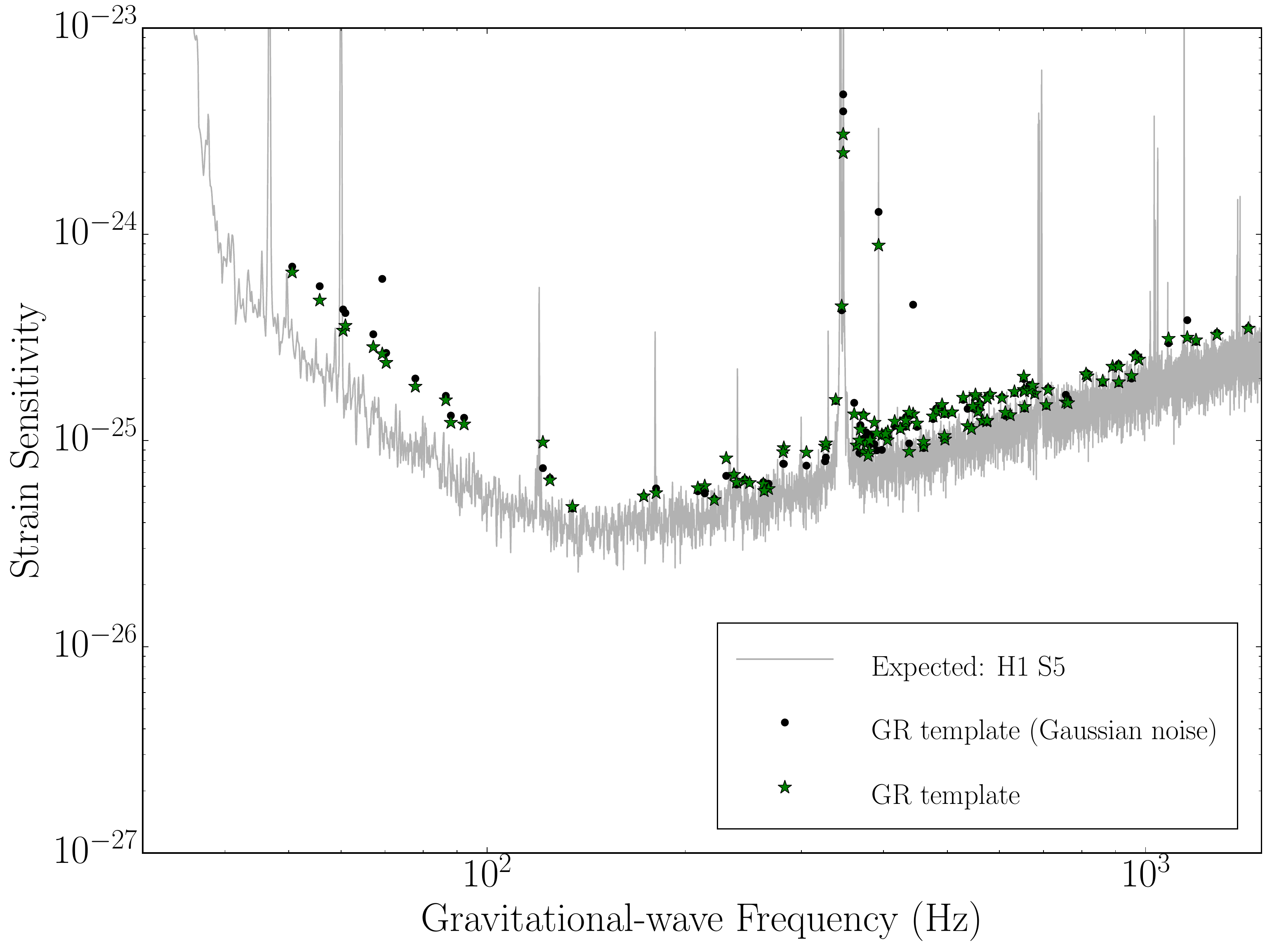}
        \caption{GR injections on Gaussian noise}
\label{fig:gaussmpGR}
\end{subfigure}%
\hfill
%\end{figure*}
\begin{subfigure}{\textwidth}
%\begin{figure*}[hbtp]
                \includegraphics[width=.7\textwidth]{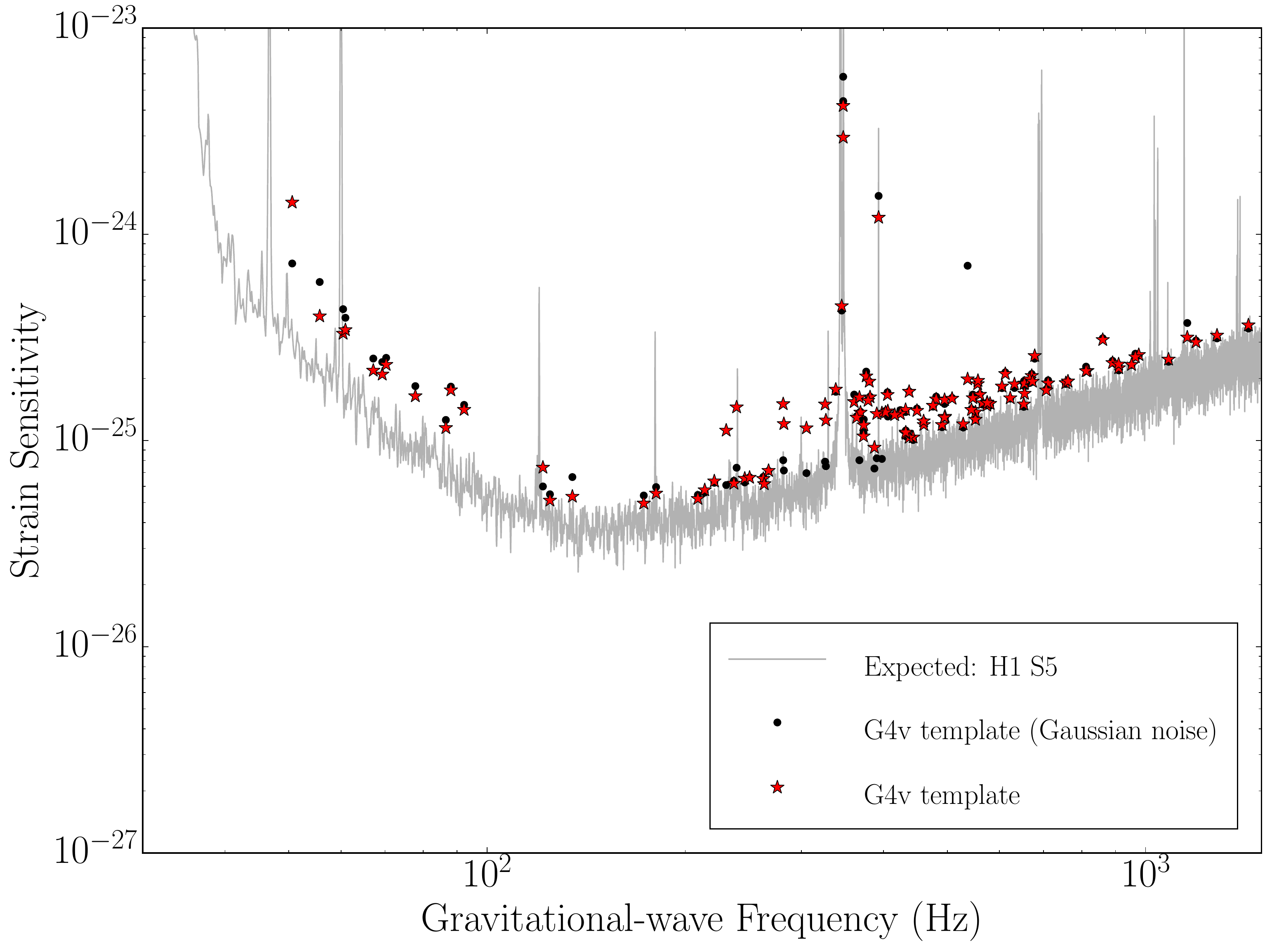}
        \caption{G4v injections on Gaussian noise}
\label{fig:gaussmpG4v}
\end{subfigure}
\caption{Expected sensitivity ($\fracdetthrsh=99.9\%,~\fracdetconf=95.0\%$) vs.~GW frequency. Comparison between fabricated Gaussian noise and actual LIGO noise. Searches were made with semi--model--dependent templates, eqs.~(\ref{eq:tempGR}, \ref{eq:tempG4v}). The colored stars correspond to actual LIGO H1 noise  (cf.~fig.~\ref{fig:mpH1}), while the black dots correspond to fabricated Gaussian noise.}
\label{fig:gaussmp}
\end{figure*}

In order to confirm that our assumption of Gaussianity is not too far from reality, we repeated our analysis (see section \ref{sec:results}) on sets of synthetic Gaussian noise. In order to do this, for each pulsar we generated streams of complex--valued data randomly selected from a normal distribution with the same standard deviation as the corresponding original LIGO data set. These series replaced the instantiations of re--heterodyned data, but the search process was otherwise unchanged. The results of this comparison for S5 H1 are shown in figs.~\ref{fig:gaussmp}, where we juxtaposed expected sensitivities obtained using Gaussian noise and actual LIGO noise (cf.~section \ref{sec:results}). These plots confirm that, indeed, we obtain qualitatively the same results with Gaussian noise as with actual LIGO data.

\bibliography{ligo}

\end{document}